%% file: sample-sigconf.tex
\documentclass[10pt,journal,compsoc]{IEEEtran}
\ifCLASSOPTIONcompsoc
  \usepackage[nocompress]{cite}
\else
  \usepackage{cite}
\fi

\ifCLASSINFOpdf
   \usepackage[pdftex]{graphicx}
   \DeclareGraphicsExtensions{.pdf,.jpeg,.png}
\else
   \usepackage[dvips]{graphicx}
   \DeclareGraphicsExtensions{.eps}
\fi

\usepackage{mathrsfs}
\usepackage{CJKutf8}
\usepackage[outdir=./]{epstopdf}
\usepackage{url}
\usepackage{makecell}
\usepackage{xcolor}
\usepackage{multirow}
\usepackage{graphicx}
\usepackage[normalem]{ulem}
\usepackage{booktabs}
\useunder{\uline}{\ul}{}

\usepackage{bbm}
\usepackage{amsmath,amsfonts}

\usepackage{amssymb}
\usepackage{pifont}% http://ctan.org/pkg/pifont
\usepackage{caption}
\usepackage{enumitem}
\usepackage{makecell}
\newcommand{\cmark}{\ding{51}}%
\newcommand{\xmark}{\ding{55}}%

\begin{document}
\begin{CJK}{UTF8}{gbsn}

% \title{Interdisciplinary Topic Path Detection for Research Proposals via A Hierarchical Attention Model}

% \title{Hierarchical Interdisciplinary Research Proposal Classification Network for Topic Path Detection}

\title{Hierarchical Interdisciplinary Topic Detection Model for Research Proposal Classification}

% keywords:multiple text; hierarchical discipline structure; interdisciplinarity; classification

% Identify the hierarchical topics for Interdisciplinary Research Proposals via a Hierarchical multi-label classification schema.

% Identify the Interdisciplinary topic path for Research Proposals via a Hierarchical multi-label classification schema.

% Hierarchical Interdisciplinary Research Proposal Classification Network for Topic path detection.

% Transformer based Hierarchical multi-label classification for interdisciplinary proposals.

%\title{How to Assign Your Proposal for Review? Interdisciplinary Topic Path Detection for Research Proposals}

% \title{Who Should Review your Proposal?
% An Interdisciplinary Research Proposal Classification Method with the Right Granularity}

% \author{Anonymous Authors}
\author{{Meng Xiao, Ziyue Qiao, Yanjie Fu, Hao Dong, Yi Du, Pengyang Wang,  \\Hui Xiong, \textit{Fellow, IEEE}, and Yuanchun Zhou}%
\IEEEcompsocitemizethanks{\IEEEcompsocthanksitem Meng Xiao, Ziyue Qiao, Hao Dong are with Computer Network Information Center, Chinese Academy of Sciences, and University of Chinese Academy of Sciences, Beijing.
E-mails: \{xiaomeng7890, ziyuejoe\}@gmail.com, donghao@cnic.cn
\IEEEcompsocthanksitem Yanjie Fu is with Department of Computer Science, University of Central Florida, Orlando. E-mail: yanjie.fu@ucf.edu
% \IEEEcompsocthanksitem   are with Computer Network Information Center, Chinese Academy of Sciences, Beijing. E-mail: \{qiaoziyue, donghao\}@cnic.cn
\IEEEcompsocthanksitem Yi Du, and Yuanchun Zhou are with Computer Network Information Center, Chinese Academy of Sciences, Beijing, and University of Chinese Academy of Sciences, Beijing, and University of Science and Technology of China, Hefei. E-mails: \{duyi, zyc\}@cnic.cn
\IEEEcompsocthanksitem Pengyang Wang is with State Key Laboratory of Internet of Things for Smart City, University of Macau, Macau. E-mail: pywang@um.edu.mo
\IEEEcompsocthanksitem Hui Xiong is with Thrust of Artificial Intelligence, The Hong Kong
University of Science and Technology (Guangzhou), Guangzhou. E-mail: xionghui@ust.hk
\IEEEcompsocthanksitem Corresponding authors: Yi Du and Yuanchun Zhou.
}% <-this % stops an unwanted space
}
% \renewcommand{\shortauthors}{ }
% \thanks{$^*$Corresponding author.}
% \affiliation{%
% \institution{\text{$^1$Computer Network Information Center, Chinese Academy of Sciences, Beijing},
% \text{$^2$University of Chinese Academy of Sciences, Beijing},
% \text{$^3$Department of Computer Science, University of Central Florida, Orlando},
% \text{$^4$University of Macau, Macau}
% \text{$^5$National Natural Science Foundation of China, Information Center, Beijing}
% }
% \country{$^{1,2}$\{shaow, qiaoziyue,donghao\}@cnic.cn, $^1$\{duyi, zyc\}@cnic.cn, $^3$yanjie.fu@ucf.edu,$^4$pywang@um.edu.mo,\\$^5$lidong@nsfc.gov.cn}
% }
% \pagestyle{plain}
\IEEEtitleabstractindextext{%
\begin{abstract}
\input{document/abstract}

\end{abstract}

% \begin{IEEEkeywords}
% Computer Society, IEEE, IEEEtran, journal, \LaTeX, paper, template.
% \end{IEEEkeywords}
}


\maketitle

\IEEEraisesectionheading{\section{Introduction}\label{sec:introduction}}
\input{document/introduction}
\input{document/preliminaries}
\input{document/method}
\input{document/application}
\input{document/experiments}
\input{document/related}

\input{document/conclusion}
\vspace{-0.3cm}
\section{Acknowledgments}
% \vspace{-0.2cm}
% This research was supported by the Natural Science Foundation of China under Grant No. 61836013, Ministry of Science and Technology Innovation Methods Special work Project under grant 2019IM020100, Beijing Nova Program of Science and Technology under Grant No. Z191100001119090, Beijing Natural Science Foundation under Grant No.4212030 and Youth Innovation Promotion Association CAS.
This work was supported by  the Natural Science Foundation of China under Grant No.61836013, the National Key Research and Development Plan of China under Grant No. 2022YFF0712200, Beijing Natural Science Foundation under Grant No.4212030, Youth Innovation Promotion Association, CAS.
\vspace{-0.3cm}
\bibliographystyle{IEEEtran}
\bibliography{bib/reference}
\input{document/bio}

% \clearpage
% \appendix
% \section{Appendix}
% \input{document/appendix}
\end{CJK}
\end{document}

%% file: document/abstract.tex
%Yanjie
The peer merit review of research proposals has been the major mechanism to decide grant awards. 
However, research proposals have become increasingly interdisciplinary. 
It has been a longstanding challenge to assign interdisciplinary proposals to appropriate reviewers so proposals are fairly evaluated. 
One of the critical steps in reviewer assignment is to generate accurate interdisciplinary topic labels for proposal-reviewer matching. 
Existing systems mainly collect topic labels  manually generated by principle investigators. However, such human-reported labels can be non-accurate, incomplete, labor intensive, and time costly. 
What role can AI play in developing a fair and precise proposal reviewer assignment system? 
In this study, we collaborate with the National Science Foundation of China to address the task of automated interdisciplinary topic path detection. 
For this purpose, we develop a deep Hierarchical Interdisciplinary Research Proposal Classification Network (HIRPCN). 
Specifically, we first propose a hierarchical transformer to extract the textual semantic information of proposals. 
We then design an interdisciplinary graph and leverage GNNs to learn representations of each discipline in order to extract interdisciplinary knowledge. 
After extracting the semantic and interdisciplinary knowledge, we design a level-wise prediction component to fuse the two types of knowledge representations and detect interdisciplinary topic paths for each proposal. 
We conduct extensive experiments and expert evaluations on three real-world datasets to demonstrate the effectiveness of our proposed model.

%% file: document/introduction.tex
\begin{figure*}
  \centering
  \captionsetup{justification=centering}
  \includegraphics[width=0.85\textwidth]{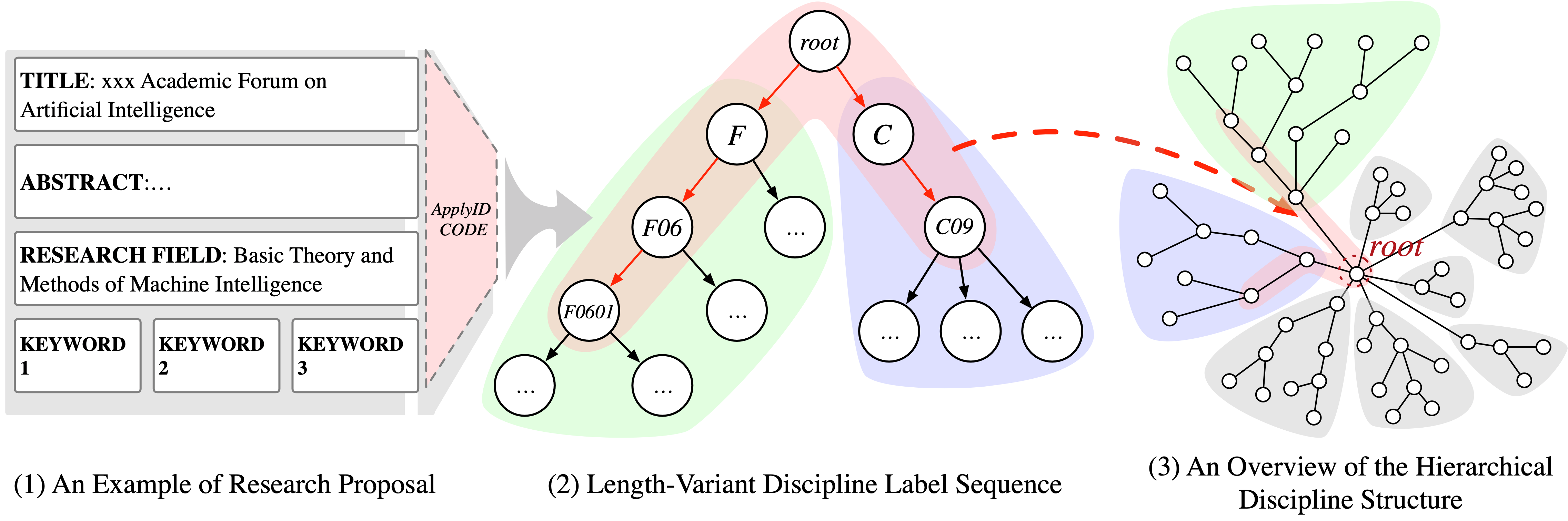}
  \vspace{-0.2cm}
  \caption{A toy model of the Interdisciplinary Research Proposal Classification.}
  \label{Fig.label_1}
  \vspace{-0.6cm}
\end{figure*}

\IEEEPARstart{N}{owadays}, research funding is awarded based on the intellectual merits and broader impacts of proposals. Under this mechanism, investigators submit research proposals to open-court competitive programs managed by federal agencies (e.g., NSF, NIH). 
Proposals are then assigned to appropriate reviewers to solicit review comments and ratings. 
The biggest challenge of running such a peer-review system is to label precise interdisciplinary topics of proposals and allocate appropriate domain experts to obtain effective review comments. 
However, with the increasing number of universities, faculty members,  hired graduate students, and publications,  the submissions of research proposals have been exploding~\cite{milojevic2015, fortunato2018science}. 
For example, the National Natural Science Foundation of China (NSFC) organizes review panels for more than one million research proposals every year. 
Due to this growth momentum, there is an urgent need to introduce AI to automated and intelligent proposal labeling for review assignment. 

In addition, research proposals increasingly exhibit interdisciplinary characteristics. 
It is inappropriate to select reviewers from only one field. 
It, therefore, is highly needed to classify a proposal into one or more disciplines, which we call the task of Interdisciplinary Research Proposal Classification (IRPC). 
Figure~\ref{Fig.label_1} demonstrates an example of the IRPC task.
After analyzing large-scale proposals from NSFC, we identify three unique data characteristics. These properties provide great potential to overcome the challenges in the IRPC task:

% 1/ 文字的传达精确 
% 2/ 图表尽量接近文字，不要出现长距离引用，方便阅读 
% 3/ an a the 
% 4/ 部分文字需要配合e.g.解释 
% 5/ 不要出现没有解释过的名词
% 6/ noun. 
(\textbf{Property 1: Hierarchical Discipline Structure})   % Structure of Academic Disciplines
% The domain knowledge, peer-reviewers, and the proposal categories are natural to organize into a discipline system. 
% Before embarking on the problem, it is worth asking how modern society regiment the scientific community including domain knowledge, peer-reviewers, and the proposal categories. 
Before embarking on the specific problem, it is worth understanding how the funding administrator regiment the disciplines. 
In NSFC,  thousands of disciplines and sub-disciplines are organized into a hierarchical structure to better index, search, and match reviewers and proposals. 
Each discipline is associated with a unique alphabetical identifier, which is called ApplyID\footnote{\url{https://www.nsfc.gov.cn/publish/portal0/xmzn/2020/16/}}. 
An ApplyID is a sequence of alphabetical letters.  
The ApplyID starts with a capital letter from \textit{A} to \textit{H}, representing eight major disciplines (e.g., \textit{Mathematical Sciences} is marked by the letter \textit{A}). 
The capital letter is followed by a sequence of numbers, in which every two numbers stand for a sub-discipline division. 
Figure~\ref{Fig.label_1}.2 shows an example disciplinary path: F0601. \textit{F} represents Information Sciences. \textit{06} represents \textit{Artificial Intelligence} that is a sub-discipline of \textit{Information Sciences}.  \textit{01} represents \textit{Fundamentals of Artificial Intelligence} that is a sub-discipline of artificial intelligence. 

(\textbf{Property 2: Interdisciplinarity}) 
Interdisciplinary research is an indispensable part of advancing scientific progress~\cite{foster2015tradition}. 
In NSFC, an interdisciplinary research proposal will be marked by two ApplyID codes and assigned to the reviewers according to those codes. 
Figure~\ref{Fig.label_1}.2 shows  the ApplyID for an interdisciplinary proposal. 
This proposal is related to  Fundamentals of Artificial Intelligence from the major discipline Information Sciences (\textit{F0601}) and Neuroscience and Psychology (\textit{C09}) from  the major discipline Life Sciences. 
A proposal with incorrect or incomplete ApplyID codes will be assigned to wrong or non-expert reviewers, which lead to biased evaluations. 
Thus, to avoid the destruction of valuable pioneering and interdisciplinary research, it is critical to identify an accurate \textbf{multiple disciplines} for interdisciplinary proposals. 

(\textbf{Property 3: Multiple-type Text}) 
The research proposal consists of multi-type textual data, including abstract, keywords, and title. Different types of text have distinct impacts on the topic path detection process. 
For example, an expert can identify the coarse-grained discipline (e.g., \textit{Information Sciences}) of a proposal by reading the title. 
However, when the expert tries to identify a fine-grained disciplinary category  (e.g., the \textit{Machine Learning}, a sub-discipline of information sciences), they might need to consider the contents of each type (e.g., the introduction).
In addition, since texts in most research proposals vary widely in length and structure, stitching each text together results in a severe loss of information.

To model the three unique proprieties, we propose three unique strategies for the IRPC task. 
First, we adopt the top-down Hierarchical Multi-label Classification (HMC) schema to utilize the \textbf{hierarchical structure} among the disciplines.
This schema organize the prediction process level-wisely, thus reducing the candidate target number. 
Further, a level-wise prediction can generate the topic path under a given partial label, which is practical for a real-world application (e.g., the funding administrators or applicants might provide an unrefined topic path, and the model can generate the rest). 
Second, the \textbf{interdisciplinarity} among the disciplines can be represented as a weighted graph. We construct an Interdisciplinary Graph and use its topology structure to learn the label (discipline) embedding, thus incorporating the knowledge into each prediction step.
Third, we model the semantic information within \textbf{multiple-type text} separately by a hierarchical Transformer architecture. In our design, each type of text will attach with a type token to preserve its type-specific semantic information. The self-attention mechanism in Transformer will also enhance the model's ability to handle the texts with variate lengths. 
Based on the above technical insights, we propose \textbf{H}ierarchical \textbf{I}nterdisciplinary \textbf{R}esearch \textbf{P}roposal \textbf{C}lassification \textbf{N}etwork (HIRPCN), a deep learning model with a hierarchical multi-label classification scheme. 

In conclusion, our contributions are as follows: 
\begin{itemize}
    \item We identify the critical problems of assigning proposals to its related areas for review. We solve these problems via detecting topic paths on the hierarchical discipline structure.
    % by using the type-specific semantic information and interdisciplinary knowledge.
    
    % proposal area assignment for review.
    
    % of reviewer-proposal assignment by finding the suitable research domian of the proposal. We solve these problems via detecting topic paths on the hierarchical discipline tree by using the semantic information and interdisciplinary knowledge.
    
    % \item We propose HIRPCN, a framework integrating proposals' semantic information and interdisciplinary knowledge for hierarchical label generation, while can initiatively detect one or more topic paths for interdisciplinary or non-interdisciplinary proposals, resepectively.
    \item We propose HIRPCN, a framework integrating proposals' semantic information and interdisciplinary topology knowledge for hierarchical label generation, which can generate one or more topic paths for non-interdisciplinary or interdisciplinary proposals. 
    % \textb{\color{red} //?? did you describe interdisciplinary knowledge before? }
    
    % for hierarchical label generation of research proposals using the type-specific semantic information and interdisciplinary knowledge, while can generate multiple topic paths for interdisciplinary proposals.
    
    \item The experiments and experts' evaluations show that our model can generate high-quality prediction results and reveal the candidate interdisciplines for research with incomplete labels.
    
    \item We post a demo\footnote{Demo available in \url{http://101.35.54.56/}} of our model to demonstrate the process of the topic path detection and how it would aid the funding administrator. Our approach will apply to funding management in the future and contribute to improving the effectiveness and fairness of the science ecosystem.
\end{itemize}
% \begin{figure}
%   \centering
%   \captionsetup{justification=centering}
%   \includegraphics[width=0.45\textwidth]{figure/demo.pdf}
%   \caption{A demo of HIRPCN to aid the online interdisciplinary research proposal classification.}
%   \label{Fig.demo}
% \end{figure}

%% file: document/preliminaries.tex
\vspace{-0.4cm}
\section{Definitions and Problem Statement}

\subsection{Important Definitions}

\subsubsection{A Research Proposal}
Applicants write proposals to apply for grants. 
Figure~\ref{Fig.label_1}.1 shows a proposal which includes multiple documents such as \textit{Title}, \textit{Abstract}, and \textit{Keywords}, \textit{Research Fields}, etc.
Let's denote a proposal by $A$, the documents in a  proposal are denoted by $D=\{d_t\}_{t=1}^{|T|}$ and the types of each document are denoted by $T =\{t_i\}_{i=1}^{|T|}$. 
The $|T|$ is the total number of the document types, and $d_i$ is the document of $i$-th type $t_i$.  
Every document $d_i$ in the proposal, denoted as $d_i=[w^{1}_i,w^{2}_i,...,w_{i}^{|d_i|}]$, is a sequence of words, where $w^{k}_i$ is the $k$-th word in the document $d_i$. 

\vspace{-0.2cm}
\subsubsection{Hierarchical Discipline Structure}\label{d1}
A hierarchical discipline structure, denoted by $\gamma$, is a DAG or a tree that is composed of discipline entities and the directed \textit{Belong-to} relation from a discipline to its sub-disciplines.
% $\mathbf{C}=(C_0, C_1,C_2,...,C_H)$ $C_i=\{c^1_i,c^2_i,...,c^{\tiny |C_i|}_i\}$
The discipline nodes set $\mathbf{C}=\{C_0\cup C_1\cup ...\cup C_H\}$ are organized in $H$ hierarchical levels, where $H$ is the depth of hierarchical level, $C_k=\{c^i_k\}^{|C_k|}_{i=1}$ is the set of the disciplines in the $i$-th level.
%and $|C_i|$ is the total number of disciplines in $C_i$.
The $C_0=\{root\}$ is the root level of $\gamma$. To describe the connection between different disciplines, we introduce $\prec$, a partial order representing the \textit{Belong-to} relationship. $\prec$ is asymmetric, anti-reflexive and transitive\cite{wu2005learning}:
\begin{small}
 \begin{align} 
   & \bullet \text{The only one greatest category }\textit{root}\text{ is the root of the } \gamma,  \nonumber\\
   % root是所有节点的起点
   & \bullet \forall c^x_i \in C_i, c^y_j \in C_j, c^x_i \prec c^y_j \to \space c^y_j \not\prec c^x_i,  \nonumber\\
   % 关系不是对称的
   & \bullet \forall c^x_i \in C_i, c^x_i \not\prec c^x_i, \nonumber\\
   %自身不包含
   & \bullet \forall c^x_i \in C_i, c^y_j \in C_j, c^z_k \in C_k, c^x_i \prec c^y_j \land c^y_j \prec c^z_k \to c^x_i \prec c^z_k.  \nonumber
   % 关系可以传递
   \nonumber
 \end{align}
\end{small} 
Finally, we define the Hierarchical Discipline Structure $\gamma$ as a partial order set $\gamma=(\mathbf{C},\prec)$.
\vspace{-0.2cm}
\subsubsection{Topic Path}\label{topic_path}
In this paper, we define the target labels of each research proposal as the Topic Path $\mathbb{L} = [L_0, L_1, L_2,..., L_{H_A}]$. $L_i=\{l_i^j\}_{j=1}^{|L_i|}$ is a set of the labels in the label paths on the $i$-th level (i.e., $\forall l_i^j \in L_i \to l_i^j \in C_i$). $H_A$ is the proper length of label paths. 
For example, in the Figure~\ref{Fig.label_1}, the input research proposal are labeled by the ApplyID codes $F0601$ and $C09$. 
So the $\mathbb{L}$ for this research proposal can be processed as $[\{l_{root}\}, \{F,C\}, \{F06, C09\}, \{F0601, l_{stop}\}]$. 
The $l_{root}$ stands for the root node of the discipline structure. 
The $l_{stop}$ is an \textbf{end-of-prediction} token, which is appended to the last set in $\mathbb{L}$ to denote the topic path ended.

\vspace{-0.2cm}
\subsubsection{Interdisciplinary Graph} 
\label{co} 
% \note{modify it, note the interaction between disciplines \& shrink the edges to weight or adj matrix}
A discipline is a combination of domain knowledge and topics. On the one hand, those topics will be carefully selected~\cite{he1999knowledge} by the applicants to represent the key idea of their research proposal. On the other hand, the overlapping topic (co-topic) between disciplines could represent their similarity. 
So, how to measure \textbf{interdisciplinarity}? S. W. Aboelela \cite{aboelela2007defining} believes that interdisciplinary research is engaging from two seemingly unrelated fields, which means: (1) these disciplines will have few overlapped topics. (2) those overlapped topics will be frequently cited by the researchers. 
In this paper, we adopt the \textit{Rao-Stirling}~\cite{stirling2007general} (RS) to measure interdisciplinarity. The RS is a non-parametric quantitative heuristic, which is widely adopted to measure the diversity in science~\cite{stirling2007general,porter2009science,rafols2010diversity},  the ecosystem~\cite{biggs2012toward} and energy security~\cite{winzer2012conceptualizing}. The RS is defined as:
\begin{equation}
\label{rs}
RS = \sum_{ab(a \not = b)} (p_a\cdot p_b)^\alpha \cdot (d_{ab})^\beta,
\end{equation}
where the first part $\left[\sum_{ab(a \not = b)}(p_a\cdot p_b)\right]$ are proportional representations of elements $a$ and $b$ in the system, and the rest $\left[\sum_{ab(a \not = b)}d_{a,b}\right]$ is the degree of difference attributed to elements $a$ and $b$. The $\alpha$ and $\beta$ are two constant to weighting the two components. 
We introduce the RS in a micro view to define the weight on $e_{a\to b}$:
\begin{equation}
\label{wab}
    e_{a \to b} = (p_{a\to b})^\alpha (d_{a\to b})^\beta,
\end{equation}
where $p_{a\to b}$ is the proportion of the proposals in discipline $a$ that contain same topics in $b$, which represents the co-selected frequency of $a$ to $b$. $d_{a\to b}$ is the proportion of topic overlap from $a$ to $b$. Their detailed definitions are as:
\begin{equation}\label{g_equation}
    p_{a\to b} = \frac{\sum_{i}^{n} \mathbbm{1}(k_i \in \{K_a \cap K_b\})F_a[k_i]}{\sum_{i}^{n}F_a[k_i]}, 
\end{equation}
\begin{equation}
    d_{a\to b} = 1 - \frac{|K_a \cap K_b|}{|K_a|},
\end{equation}
where the $\mathbbm{1}(\cdot) \to \{0, 1\}$ is the indicator function, $K_a$ and $K_b$ are topic set of $a$ and $b$, $n$ is the total number of the topic. $F_a$ and $F_b$ are two lookup table holding the frequency of each topic being selected in the proposals that relat to $a$ and $b$. 

By that, we build a \textit{Interdisciplinary Graph}, denoted as $G=(\mathbf{C}, E)$, to represent the interdisciplinarity among disciplines. $G$ is a collection of discipline set $\mathbf{C}$ and directed weighted edge set $E=\left\{e_{a\to b}\right\}^{|\mathbf{C}|}_{a,b=1}$. Each $e_{a\to b} \ge 0$ represents the interdisciplinary strength from discipline $a$ to discipline $b$.
% In our paper, we assume the two components are equally important and set the $\alpha$ and $\beta$ in Equation~\ref{wab} as 1.

% \subsubsection{Interdisciplinary Distance}
% {\color{blue}}
\vspace{-0.4cm}
\subsection{Problem Formulation}
% Finally, we provide a formal definition of IRPC task. Given a proposal $A$, we aim to extract its semantic information from the document set $D$ and incorporate the information from the interdisciplinary graph $G$, and assign several label paths for it, where each label path belongs to one of the paths on the hierarchical discipline structure $\gamma$ started from the root node with arbitrary length and each label is the discipline on the paths. Figure 1 demonstrates a toy model of the IRPC problem.

We formulate the IRPC task in a hierarchical classification schema and use a sequence of discipline-level-specific label sets $\mathbb{L}$ as topic path to represent the proposals' disciplinary codes. 
To sum up, given the proposal's document set $D$ and the interdisciplinary graph $G$,
we decompose the prediction process into an top-down fashion from the beginning level to the certain level on the hierarchical discipline structure $\gamma$.
Suppose the $k-1$ ancestors labels in $\mathbb{L}$ are $\mathbb{L}_{<k} = [L_{0}, L_1,..., L_{k-1}]$, where $\mathbb{L}_{<1}=\{L_0\}$.
The prediction on level $k$ can be seen as a multi-label classification task on $C_k$ based on the incomplete topic path $\mathbb{L}_{<k}$, this process can be defined as:
$$\Omega(D,G,\gamma,\mathbb{L}_{<k};\Theta) \to L_{k},$$ 
where $\Theta$ are the parameters of model $\Omega$.  
Also, we use the \textbf{end-of-prediction} label $l_{stop}$ in the last set $L_{H_A}$ to mark the proper level (i.e., the $H_A$) where the model to stop on. 
Eventually, we can formulate the probability of the topic path assignment as:
\begin{equation}
    P(\mathbb{L}|D,G,\gamma;\Theta)=\prod_{k=1}^{H_A} P(L_k|D,G,\gamma,\mathbb{L}_{<k}; \Theta)
    \label{equation_1}
\end{equation}
where $P(\mathbb{L}|D,G,\gamma;\Theta)$ is the overall probability of the proposal $A$ belonging to the label set sequence $\mathbb{L}$, $P(L_k|D,G,\gamma,\mathbb{L}_{<k}; \Theta)$ is the label set assignment probability of $A$ in level-$k$ when given the previous ancestor $\mathbb{L}_{<k}$.
In training, given all the ground truth labels, our goal is to maximize the Equation \ref{equation_1}.

%% file: document/method.tex
\vspace{-0.4cm}
\section{Interdisciplinary Topic Detection}

\begin{figure}
\centering
\captionsetup{justification=centering}
\includegraphics[width=0.45\textwidth]{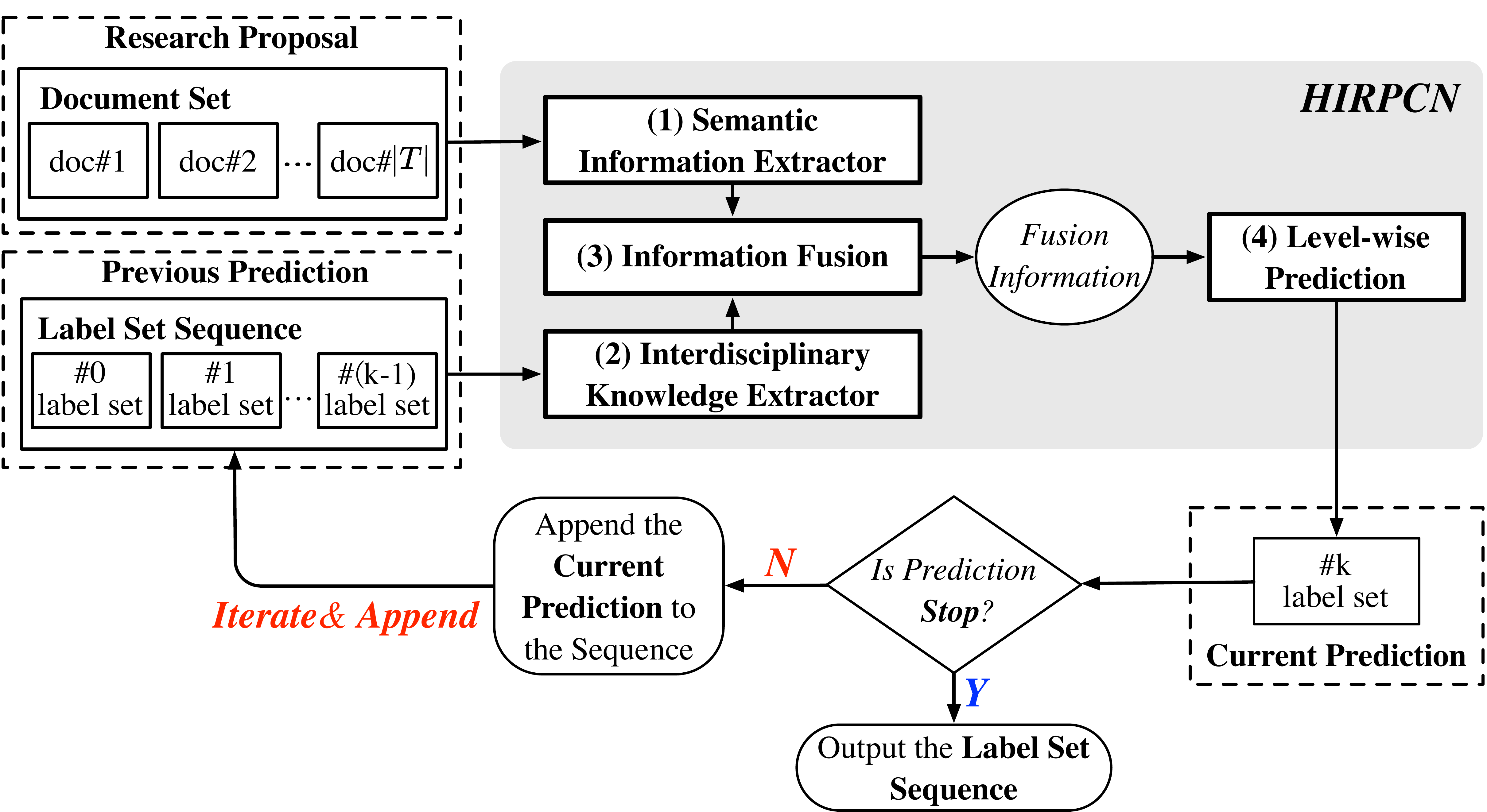} 
\caption{An overview figure of the HIRPCN during the level-$k$ prediction.}
\label{Fig.model_1}
\vspace{-0.6cm}
\end{figure}
\subsection{Overview of the Proposed Framework} Figure~\ref{Fig.model_1} illustrate iteration of our  framework upon four components: 
(1) \textit{Semantic Information Extractor} (SIE) models the semantic components in research proposals according to their particular type (e.g., title, abstract, keywords, research fields) and encodes them into a matrix. 
(2) \textit{Interdisciplinary Knowledge Extractor} (IKE) obtains the representation of previous prediction results from the pre-built interdisciplinary graphs. 
(3) \textit{Information Fusion} (IF) fuses the type-specific semantic information and the representation of previous prediction results. 
(4) \textit{Level-wise Prediction} (LP) predicts the probability that a research proposal falls within particular disciplines at the current step. 
After completing this iteration, our framework takes the discipline predictions, appends them to the label set sequence, and then moves to the next step.
The initial input of the model will be the \textbf{root node} in the hierarchical discipline structure and a document set from a research proposal. 
The model will incrementally complete the input research proposal's label path as the iteration goes on until the \textbf{end-of-prediction label} has been generated.

\begin{figure}
\centering
\captionsetup{justification=centering}
\includegraphics[width=0.47\textwidth]{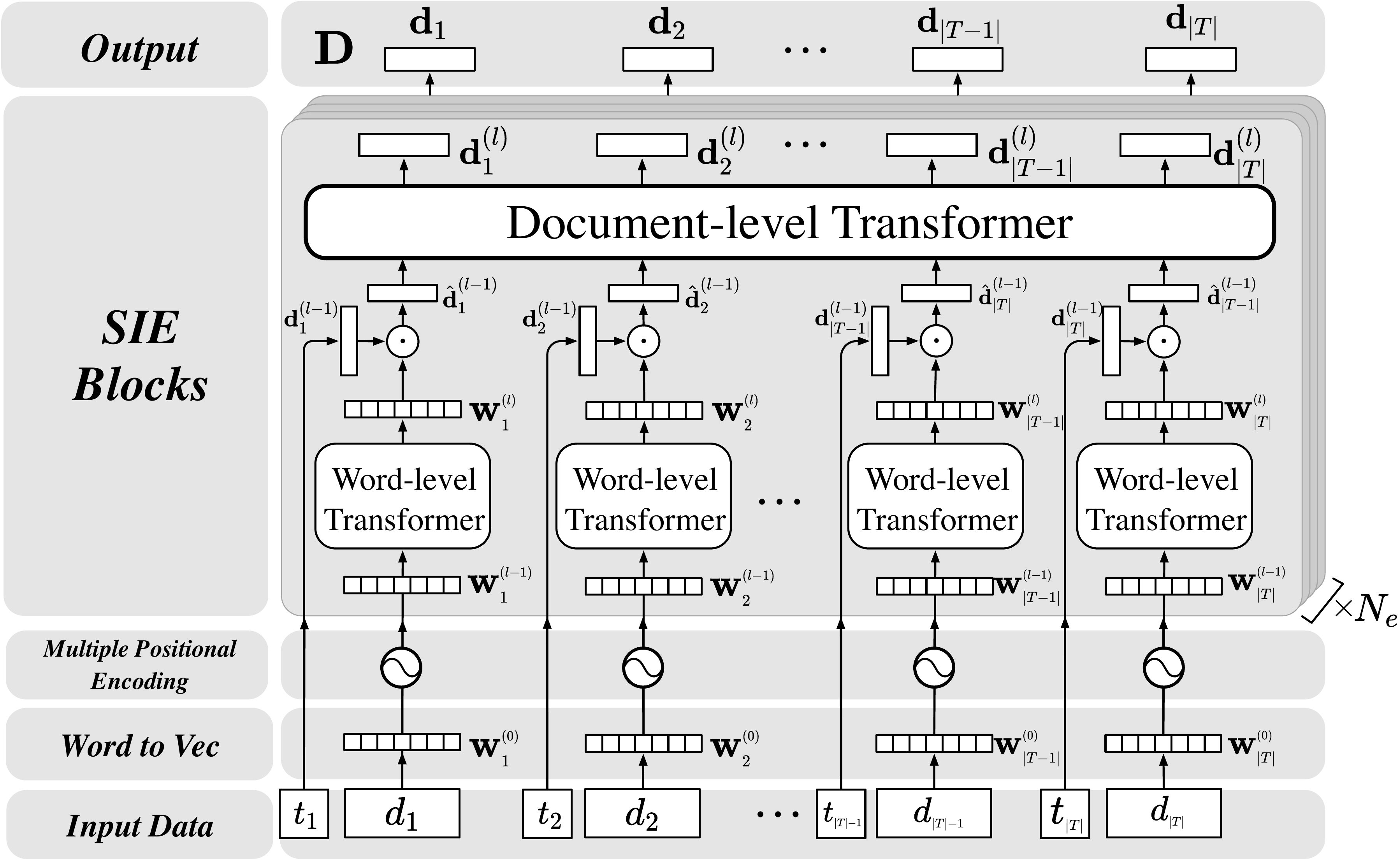}
\caption{Illustration of Semantic Information Extractor. The $\odot$ means integrating operation between vanilla document representation and their type information.}
\label{Fig.model_encoder}
\vspace{-0.5cm}
\end{figure}
\vspace{-0.45cm}
\subsection{Semantic Information Extractor}\label{SIE}
\vspace{-0.1cm}
Figure~\ref{Fig.model_encoder} shows the overview of SIE. 
SIE comprises a \textit{Word to Vec} layer, a \textit{Multiple Positional Encoding} (MPE) layer, and multi-layer \textit{SIEBlocks}. Each SIEBlock consists of two parts:  a word-level Transformer\footnote{The detailed explanation of the Transformer can be found in \cite{vaswani2017attention}.} and a document-level Transformer. Formally, the representation of a proposal is obtained by:
\begin{equation}
    \mathbf{D} = N_e \times SIEBlock(\{\mathbf{W}_i^{(0)}\}_{i=1}^{|T|}),
\end{equation}
where the matrix $\mathbf{D}$ is the representation of the input research proposal. 
$\mathbf{W}_i^{(0)}$ is the word embedding matrix of the document $d_i$. The $N_e$ is the total layer number of SIEBlocks.
% \note{represents that?}

% \vspace{0.1cm}
\noindent\textbf{Our Perspective:} We consider two essential aspects of the research proposal. 
First, different types of documents (e.g., title, abstract, keywords, research fields)  in a research proposal should be separately weighed when identifying the topic path of the proposal.
Second, in the proposal, each document varies in length (e.g., title and abstract), and concatenating them together will cause a severe loss of information. 
Inspired by the long-text modeling methods~\cite{pappagari2019hierarchical,zhang2019hibert,liu2019hierarchical}, we design a \textit{hierarchical Transformer architecture}  to embed both texts and their particular type within a proposal into a matrix. 
% \note{the hidden?}
% \note{the shape of Matrix and vector?}

\noindent\textbf{Step 1: Word to Vec and Multilple Positional Encoding.} 
We first vectorize each words in document by a pre-trained $h$-dimensional Word2Vec~\cite{mikolov2013distributed} model, then sum them with positional encoding to preserve the position information. After separately concatenating each document's word vector,  we can obtain the word embedding. It is worth noting that we padded each document to an identical length before the Word to Vec component. The length of each document after padding is denoted by $|d|$. The word embedding for document $d_i$ is denoted as $\mathbf{W}^{(0)}_i\in \mathbb{R}^{|d| \times h}$. Then we seperately feed each document representation matrix into the multi-layer SIEBlocks.
 
\noindent\textbf{Step 2: Word-level Transformer.}
In the first stage of $l$-th SIEBlock, we aim to use a word-level Transformer to extract the contextual semantic information. We can obtain the document representation matrix $\mathbf{W}^{(l-1)}_{i}  \in \mathbb{R}^{|d| \times h}$ from the previous layer. Then, we can formulate the process of the current layer as:
\begin{equation}
\begin{aligned}
    \mathbf{W}^{(l)}_{i} &= Transformer_{w}(\mathbf{W}^{(l-1)}_i),
\end{aligned}
\end{equation}
where the $\mathbf{W}^{(l)}_i \in \mathbb{R}^{|d| \times h}$ is the \textit{vanilla document representation matrix} in current layer. The ${Transformer}_{w}(\cdot)$ is the word-level Transformer block in $l$-th SIEBlock.
 
% \noindent\textbf{Step 3: Document-level Transformer.}
\noindent\textbf{Step 3: Document-level Transformer.}
In the second stage of $l$-th SIEBlock, we aim to fuse each vanilla document representation matrix with its correlated type information by a document-level Transformer:
\begin{equation}
    \begin{aligned}
        \label{trans2}
        \mathbf{D}^{(l)} = {Transformer}_{d}(\{\mathbf{d}^{(l-1)}_i \odot \mathbf{W}^{(l)}_{i}\}_{i=1}^{|T|}), 
    \end{aligned}
\end{equation}
where ${Transformer}_{d}(\cdot)$ is the document-level Transformer block. $\mathbf{d}^{(l-1)}_i  \in \mathbb{R}^{h}$ is the type-specific representation of $d_i$ from previous layer.In the first layer, the model will initialize each document's type-specific representation with a randomly initialized look-up table according to their type $t$. 
And the $\odot$ is a fusing operation. 

Inspired by ViT\cite{dosovitskiy2020image} and TNT\cite{han2021transformer}, we set the $\odot$ operation as:
(1) a \textit{Vectorization Operation} on $\mathbf{W}^{(l)}_i$.
(2) a \textit{Fully-connected Layer} to transform the vanilla document representation from dimension $|d|h$ to dimension $h$. 
(3) an \textit{Element-wise Add} with the vector $\mathbf{d}^{(l-1)}_i$ from previous layer to obatin the type-specific representation in current layer.
$\mathbf{D}^{(l)}\in \mathbb{R}^{|T|\times h}$ is the outputs of current layer, which can be seen as $|T|$-views of high-level abstraction for proposals.  
After $N_e$ times propagation, we can obtain the final output of SIE as $\mathbf{D} = \mathbf{D}^{(N_e)}$. 

\begin{figure}
\centering
\captionsetup{justification=centering}
\includegraphics[width=0.44\textwidth]{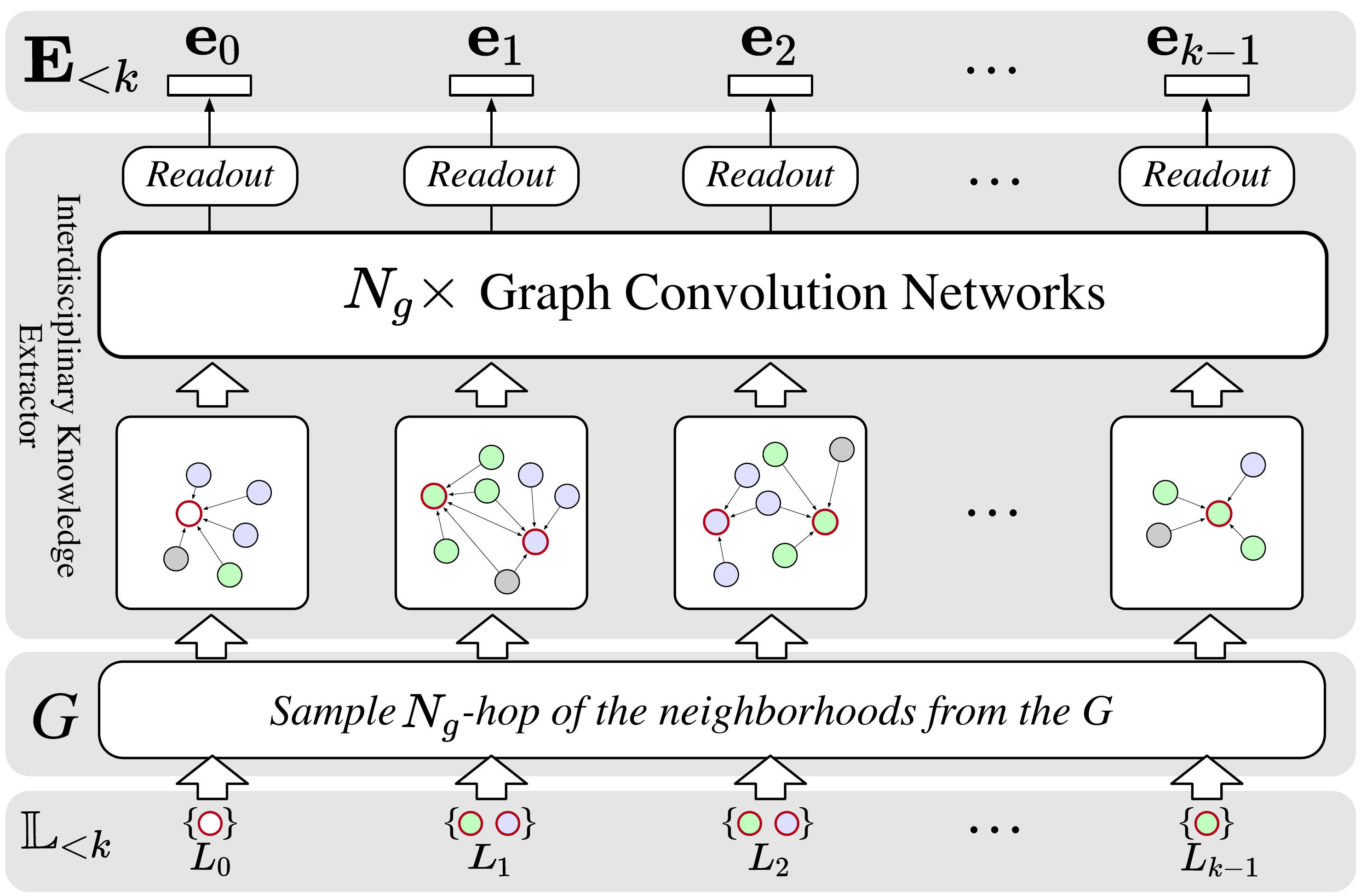} 
% \vspace{-1cm}
\caption{Illustration of Interdisciplinary Knowledge Extractor. The colors of the circles represent their belonging disciplines. We use the red outline to mark the central nodes.}
% \note{1.central nodes 2.circles}
\label{Fig.model_encoder_2}
\vspace{-0.55cm}
\end{figure}

\vspace{-0.35cm}
\subsection{Interdisciplinary Knowledge Extractor}
\vspace{-0.05cm}
The IKE is aimed to acquire the embedding of each prediction step from the pre-built interdisciplinary graph $G$. 
Figure~\ref{Fig.model_encoder_2} shows an overview of IKE, which consists of multi-layer \textit{Graph Covolutional Network}\footnote{The detailed explanation of the Graph Covolutional Network can be found in \cite{kipf2016semi}} (GCN) and a \textit{Readout Layer}. From an overall perspective, the label embedding is learned by:
\begin{equation}
    \mathbf{E}_{<k} = IKE(\mathbb{L}_{<k}, G),
\end{equation}
% \note{qzy: added $G$}
where the $\mathbf{E}_{<k} \in \mathbb{R}^{k\times h}$ is the matrix to preserve the prediction state and interdisciplinary knowledge from previous $k-1$ steps. 

\noindent\textbf{Our Perspective: Embed the Previous Prediction.}
We usually model people's behavior by capturing their relationships with others on the social network graph. The same idea is adopted in IKE to extract the previous prediction state from the interdisciplinary graph.  % 主谓宾
Instead of inputting the former prediction result directly or passing a hidden state between each step, our design uses the previous results and their neighborhood on the interdisciplinary graph to obtain the representation of each step. 
% In IKE, our fundamental idea is to learn the disciplines embedding from its interdisciplinary interaction with other disciplines.\note{qzy:to discuss. The idea is use graph to model and extract interdisciplinary information} 
% To utilize the knowledge, we first model the interdisciplinary relation as $G$, then use Graph Convolutional Networks (GCNs) to aggregate the neighborhood information into the predicted label embedding. 
% 将交叉学科互动融入学科表征
% 我们的观点：Our Perspective: Modeling interdisciplinary interactions for label embeddings. 为了使label embedding获得交叉学科信息。我们首先将交叉关系建模为graph。接着我们使用GNN将有交叉关系的学科聚合到中心学科上，使其获得交叉学科信息。

% Our solution is to represent the predicted label with its sub-graph (to describe its behavior) between each step.

\noindent\textbf{Step 1: Obtain the Representation.} 
Suppose we are processing the $k$-th step. 
Given the sequence of label sets $\mathbb{L}_{<k}= [L_{0}, L_1,..., L_{k-1}]$ from the previous $k-1$ steps (or from a manually given partial label).
Each label set $L_i$ consists of the disciplines for holding the prediction result in the $i$-th step. 
We first treat the including discipline nodes as the \textbf{central node} and sample their $N_g$-hops neighborhoods (as an adjacency weighted matrix) from the interdisciplinary graph $G$, where the $N_g$ is the total layer number of GCN. 
Then we feed the adjacency weighted matrix and its corresponding node features into the GCN\footnote{We implemented the GCN with the message-passing design~\cite{wang2019dgl}.}:
\begin{equation}
    \mathbf{H}_i = N_g \times GCN(E_i, \mathbf{H}^{(0)}_i),
\end{equation}
where the $\mathbf{H}^{(0)}_i$ is a random initialized input feature of the central nodes and their $N_g$-hops neighborhood nodes. The $E_i$ is the weighted adjacency matrix between these nodes. The $\mathbf{H}_i$ is the output matrix undergoing $N_g$ times GCN propagations. 

\noindent\textbf{Step 2: Readout the Features.} 
% \note{QZY: h represents a matrix, should be H. It doesn't feel necessary to divide into two steps.
% If two steps, i suggest step 1 is subgraph sampling and step 2 is subgraph encoding.
% }
Due to the varied number of discipline nodes within each prediction label set, we have to pool the matrix and acquire the representations of each step.
For example, the representation of $L_i$ is obtained by:
\begin{equation}
    \mathbf{e}_i = Readout(\mathbf{H}_i, L_i),
\end{equation}
where the \textit{Readout} layer is a lookup operation (i.e., take the $L_i$ included node embedding) followed by a mean pooling layer. 
The $\mathbf{e}_i \in \mathbb{R}^h$ is the final representation of $L_i$. Then we can obtain the output of IKE as $\mathbf{E}_{<k}=[\mathbf{e}_0, \mathbf{e}_1,...,\mathbf{e}_{k-1}]$. 

The $\mathbf{E}_{<k}$ can be seen as the representation of the prediction results from the previous $k-1$ steps integrated with their neighborhoods information from the interdisciplinary graph.
We believe that the neighborhood information could help the model more inclined to predict the discipline that is highly connected to the historical predicted discipline on the interdisciplinary graph, thereby improving the ability to predict the interdisciplinary.  
% We believe that the IKE component can help the HIRPCN detect the \textbf{multiple paths} for the IRPC task by utilizing each nodes' neighborhood information. 
% 对交叉学科知识的融合使得模型在下一层预测时更倾向于预测于上层学科联系紧密的学科，从而提高了预测的能力。
% multiple paths --> detect more disciplinary
% label + interdiscplinary knowledge => the knowledge help the proposal with 具有交叉性的先前预测结果，能够在下游预测中预测出与之相关的交叉性的子学科
% \note{qzy: To discuss. Can we go deeper?} 
%GNN 的propagation 使得label embedding 融合了邻域信息，进一步使得模型能够使用上层label的邻域信息做预测，有助于.
\vspace{-0.4cm}
\subsection{Information Fusion}
\vspace{-0.1cm}
\begin{figure}
\centering
\captionsetup{justification=centering}
\includegraphics[width=0.45\textwidth]{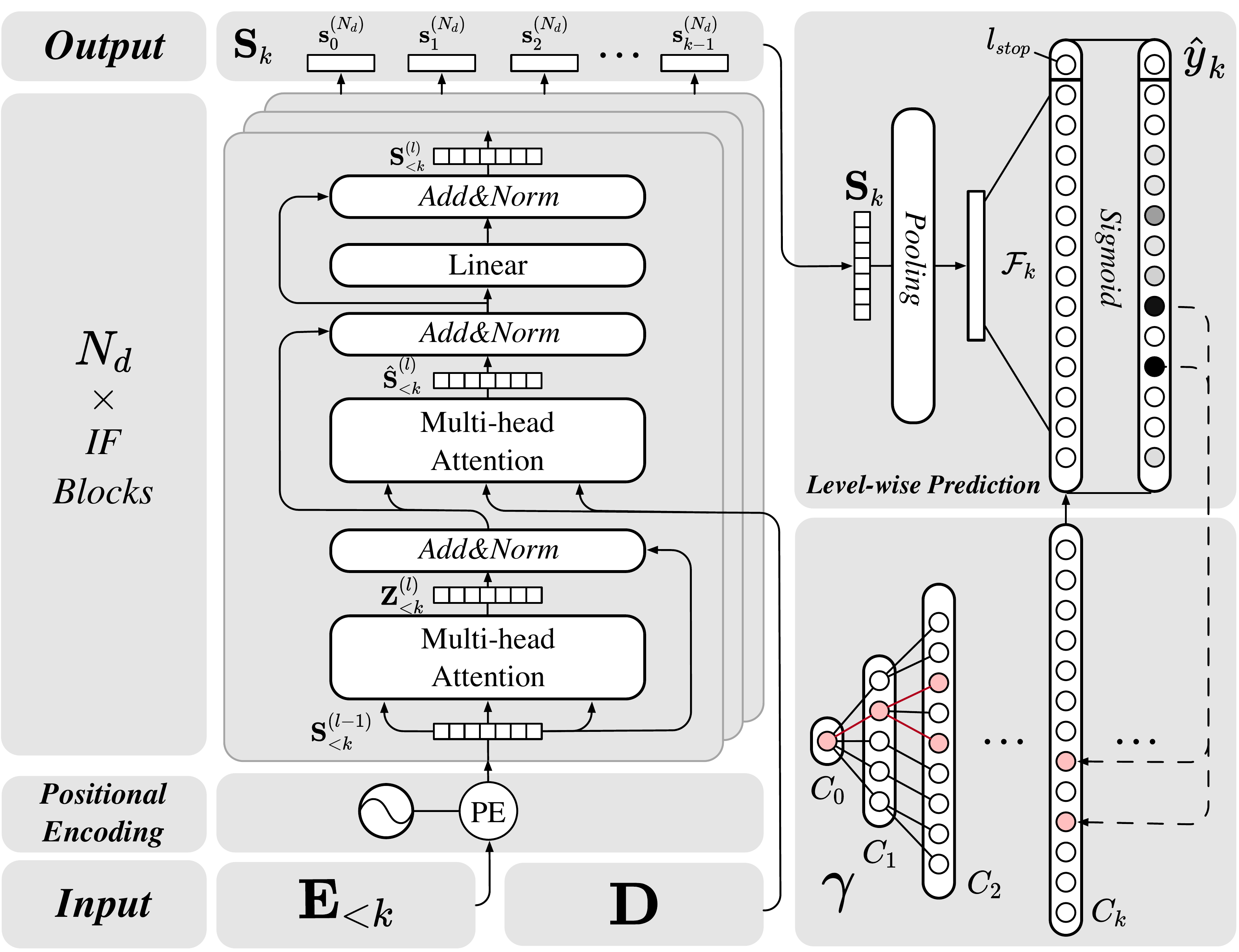}
\caption{Illustration of Information Fusion and Level-wise Prediction in step-$k$.}
\label{Fig.model_decoder}
\vspace{-0.6cm}
\end{figure}
In IF, we use a multi-layer \textit{IF Block} to integrate the extracted information from SIE and IKE. As the left side of  Figure~\ref{Fig.model_decoder} shows, the IF consists of a positional encoding and multi-layer IFBlocks. We start the introduction by giving its dataflow:
\begin{equation}
    \mathbf{S}_{k} = N_d \times IFBlock(\mathbf{E}_{<k}, \mathbf{D}),
\end{equation}
where $N_d$ is the total layer number of IF Block, and $\mathbf{S}_{k} \in \mathbb{R}^{k\times h}$ is the fusion matrix that integrated the semantic information and the previous prediction information with interdisciplinary knowledge.

\noindent\textbf{Our Perspective: Adaptively Utilize Each Part of Information.} There are two aspects we consider in IF. First, the current prediction state should be significantly affected by its previous results. Second, the model should adaptively choose the critical part of semantic information according to its current prediction state. 
To advance those, we divide the IF into two steps.    

\noindent\textbf{Step 1: Positional Encoding and Previous Label Embedding Fusion.} 
%\note{Incorporate the Label Embedding and Hierarchical Dependency.}
We first sum the $\mathbf{E}_{<k}$ with positional encoding to obtain $\mathbf{S}_{<k}^{(0)}$ for preserving the order of prediction. Then, in layer-{$(l)$} we perform \textit{Multi-Head Self-Attention} 
between each prediction result representation to help each element  in $\mathbf{S}_{<k}^{(l-1)}$ aggregate knowledge from their context: 
\begin{equation}
    \label{M}
    \hat{\mathbf{S}}^{(l)}_{<k} =  \mathbf{S}_{<k}^{(l-1)}\circ \textit{MultiHead}(\mathbf{S}_{<k}^{(l-1)},\mathbf{S}_{<k}^{(l-1)},\mathbf{S}_{<k}^{(l-1)})),
\end{equation}
where  $\circ$ operation is a \textit{Layer Normalization} with a \textit{Residual Connection Layer}.
In this step, the $\textit{MultiHead}(\cdot)$ adaptively aggregate the context information from every step prediction, and the $\circ$ operation integrate the context knowledge to the prediction result embedding and form the the level-$l$ prediction state as $\hat{\mathbf{S}}^{(l)}_{<k} \in \mathbb{R}^{k\times h}$. 
With the order information, we believe step 1 of IF can capture the hierarchical dependency and the interdisciplinary knowledge in $\mathbb{L}_{<k}$

\noindent\textbf{Step 2: Obtain the Semantic Informaion Adaptively.}
As mentioned before, we hope our model can utilize each part of the research proposal adaptively through prediction progress. Thus, we treat the prediction state $\hat{\mathbf{S}}^{(l)}_{<k}$ as \textit{Query} and the representation of document set  $\mathbf{D}$ as \textit{Key} and \textit{Value} into another \textit{Multi-head Attention} to propagate the semantic information:  
\begin{equation}
    \label{Z}
    \begin{aligned}
        \mathbf{Z}^{(l)} &= \hat{\mathbf{S}}^{(l)}_{<k} \circ \textit{MultiHead}(\hat{\mathbf{S}}^{(l)}_{<k} , \mathbf{D}, \mathbf{D})), \\
        \mathbf{S}_{<k}^{(l)} &= \mathbf{Z}^{(l)} \circ FC(\mathbf{Z}^{(l)}),
    \end{aligned}
\end{equation}
The $\textit{MultiHead}(\cdot)$ adaptively aggregate the semantic information from the research proposal by the current prediction state, and the $\circ$ operation integrate the semantic information to the label embedding and construct the hidden feature $\mathbf{Z}^{(l)} \in \mathbb{R}^{k\times h}$ in level-$l$. After a \textit{Fully-connected Layer} denoted as $FC(\cdot)$ and the $\circ$ operation, we form $\mathbf{S}_{<k}^{(l)}$ to hold the fusion information in level-$l$.  

After $N_d$ times propagations, we acquire the $\mathbf{S}_{k} = \mathbf{S}_{<k}^{(N_d)}$ to represent the output of IF. With the IF, the model will learn the dependency between previous prediction steps and adaptively fuse each part of semantic information by the current prediction progress.
\vspace{-0.3cm}
\subsection{Level-wise Prediction}
\vspace{-0.1cm}
% According to Equation \ref{equation_1}, we hope that the HIRPCN can generate labels set in a \textbf{top-down fashion} from the beginning level to a \textbf{proper} level $H_A$. 
The right side of Figure \ref{Fig.model_decoder} demonstrate the prediction in step-$k$. In LP, We feed the fusion feature matrix $\mathbf{S}_{k}$ into a \textit{Pooling Layer}, a \textit{Fully-connected Layer}, and a \textit{Sigmoid Layer} to generate each label's probability for $k$-th level-wise label prediction. 
\begin{equation}
    y_{k} = Sigmoid(FC_k(Pooling(\mathbf{S}_{k}))),
\end{equation}
where the $y_{k}$ is the probability (i.e., $P(L_k|D,G,\gamma,\mathbb{L}_{<k}; \Theta)$ in Equation~\ref{equation_1}) for each discipline in $k$-th level of hierarchical discipline structure $\gamma$. 
It is worth noting that we add a $l_{stop}$ token in each step to determine the prediction state. 
Thus, the $FC_k(\cdot)$ denotes a level-specific feed-forward network with a ReLU activation function to project the input to a $|C_k|+1$ length vector, where represents the label number of level-$k$ with $l_{stop}$ token. 
After the $Sigmoid(\cdot)$, the final output $y_k$ is the probability of $k$-th level's labels. 
Thus, the level-$k$'s objective function $\mathcal{L}_k$ can be defined as:
\begin{equation}\label{objective_function}
\begin{aligned}
    \mathcal{L}_k(\Theta) &= \sum^{\mid C_k\mid + 1}_{i=1} \left(Y_k(i) \log(y^i_k) + (1 - Y_k(i)) \log(1 - y^i_k) \right),\\
    &\text{where } Y_k(i) = \begin{cases}
0 : l_i \not\in L_k
 \\ 1: l_i \in L_k
\end{cases},
\end{aligned}
\end{equation} 
where the $y^i_k$ is the probability of the level-$k$'s $i$-th label.

The label prediction starts from the 1st level with given $\mathbb{L}_{<1}$.  
In the $k$-th level prediction, the label corresponding to the index of the value in the $y_k$ which achieves the threshold will be selected to construct the current prediction result set $L_k$.
The model will append the $L_k$ to $\mathbb{L}_{<k}$ to construct $\mathbb{L}_{k}$ as current prediction result. 
If the prediction continues,  $\mathbb{L}_{k}$ will form the next previous prediction result $\mathbb{L}_{<(k+1)}$.
To require that the model can define a distribution over labels of all possible lengths, we add a \textbf{end-of-prediction label} $l_{stop}$ token and set the first element of $y_k$ as the probability of $l_{stop}$ in level-k.  
In the final step, the $l_{stop}$ will include in the prediction target. 

We sum the loss in each level to obtain the overall loss function of HIRPCN as:
\begin{equation}\label{overall_loss}
% \begin{aligned}
        \mathcal{L}(\Theta) = \sum\mathcal{L}_k(\Theta).\\
% \end{aligned}
\end{equation} 
During the training process, our target is to optimize the objective function $\mathcal{L}(\Theta)$.
%
% Finally, after the prediction from level-$1$ to level-$H_A$, the overall objective function for each proposal $A$ with document set $D$ is to maximize the log-likelihood of the overall probability:
% % \begin{equation}
% %         \Theta = \mathop{\arg\max}_{\Theta}\sum_{k=1}^{H_A} \log P(L_k|D,\gamma,G,\mathbb{L}_{<k}; \Theta)
% % \end{equation}
% \begin{equation}
% \begin{aligned}
%     \mathcal{L}_k(\Theta)    &= \log(P(L_k|D,\gamma,G,\mathbb{L}_{<k}; \Theta))\\
%      &= \sum^{|C_k| + 1}_{i=1} \left(y_i \log(\hat{y}_k^i) + (1 - y_i) \log(1 - \hat{y}_k^i) \right) \\
% \end{aligned}
% \end{equation}
% where the $L_k$ is the ground truth label set of level-$k$ prediction. And $y_i=\mathbbm{1}(l_i \in L_k)$, which aims to discriminate whether the $i$-th label $l_i$ belongs to the truth label set or not.  

%% file: document/application.tex
\vspace{-0.3cm}
\subsection{Application}
\vspace{-0.1cm}
Our proposed method (HIRPCN) aims to generate the topic path for an input research proposal. 
In real world scenario, the applicants will submit the research proposals and the ApplyID codes. Then the fund administrator will assign reviewers depending on the given ApplyID codes. 
This framework can introduce AI as an assistant to address: 1) the barrier between the applicant and the complex discipline structure; 2) the missed or concealed topic information. 
Besides these, the design of IKE can let the model start prediction from any given level, making the applicants only need to provide coarse-grained labels, which is essential for online guidance.  
Lastly, the attention mechanism can provide interpretability for model prediction, and its visualization can facilitate the fund administrator to verify each research proposal's topic path. 

% This framework is applicable to a variety of domain tasks without the need for prior knowledge.
% We apply GRFG to classification, regression, and outlier detection tasks on 29 different datasets to validate its performance.

%% file: document/experiments.tex
\vspace{-0.4cm}
\section{Experiments}
\vspace{-0.1cm}
% \subsection{From the Perspective of Applications}
% To address a real-world problem, we should first illustrate our consideration of experiment design from the application's perspective. 
In this section, we conducted experiments to evaluate the performance of HIRPCN and answer the following questions: \textbf{Q1.} Is the performance of HIRPCN superior to the existing baseline models? 
\textbf{Q2.} How is each component of HIRPCN affect the performance? 
\textbf{Q3.} Can HIRPCN achieve the best performance on the prediction at all levels?
\textbf{Q4.} How good is the prediction result?
\textbf{Q5.} Can the given partial expert advice improve the performance of HIRPCN? 
\textbf{Q6.} Can HIRPCN adaptively utilize the semantic information and historical prediction results?
\textbf{Q7.} How is the word-level attention mechanism works for a better semantic extraction? 
\textbf{Q8.} What is the extra generated label significance to the peer-review assignment?
We also evaluated the model performance from different settings of interdisciplinary graphs and hyperparameters. 
\vspace{-0.3cm}
\subsection{Experimental Setup}
\vspace{-0.1cm}
\subsubsection{Dataset Description}
\vspace{-0.1cm}
We collected research proposals written by the scientists from 2020's NSFC research funding application platform, containing 280683  records with 2494 ApplyID codes. 
As shown in Table~ \ref{disciplines}, when the level goes deeper, since there is more number of disciplines, the classification becomes harder. 
In data processing, we chose four parts of a research proposal as the document data: 1) Title, 2) Keywords, 3) Abstract, 4) Research Field. 
The Abstract part is in long-text form, and the average length is 100. The rest three documents are all short text. 
We removed all the punctuation and padded the length of each document to 200. 
Then we added a type-token at the beginning of each document to mark its particular type (just like the [\textit{CLS}] token in the Bert\cite{devlin2018bert}).
For the label construction, we generate the topic path of each research proposal  (as described in Section~\ref{topic_path}). 
% After data pre-processing, we noticed that 44\% of the research proposal is interdisciplinary. 
% The 16\% of those interdisciplinary research contains two major discipline labels (e.g., a Bioinformatics research will be marked by two major disciplines labels, says Information Sciences and Life Science). 
% And the other 84\% include one major discipline but show interdisciplinarity in its sub-discipline (e.g., a Machine Intelligence research will be marked by one major discipline label, Information Sciences, but its sub-discipline should include  Automation Science and Artificial Intelligence).
% The rest research proposals are marked as non-interdisciplinary research by their applicants. 
We further organized and divided those processed data by their interdisciplinarity into three datasets:
\begin{itemize}
    \item \textbf{RP-all:} An overall dataset with all kinds of research proposals, whether interdisciplinary research or non-interdisciplinary research. 
    \item \textbf{RP-bi:} Including all research proposals with two major disciplines or one major discipline but gets interdisciplinarity in its sub-discipline. 
    \item \textbf{RP-differ:} Only includes the interdisciplinary research proposals with two major disciplines and its data shows the most interdisciplinarity among the three datasets.
\end{itemize}

In Table~\ref{dataset}, we counted the lengths of each label sequence and the average number of level-wise labels on three datasets. Generally speaking, from RP-all to RP-differ, the data exhibit interdisciplinarity increasingly and becomes more challenging to detect its topic path. The comparison of model performance between these datasets can evaluate the specific design of HIRPCN. 

\begin{table}
\centering
\captionsetup{justification=centering}
\caption{The discipline and sub-discipline numbers on each level of hierarchical discipline structure. }
\label{disciplines}
\vspace{-0.2cm}
\input{document/table/disciplines}
\vspace{-0.3cm}
\end{table}

\begin{table}
\centering
\captionsetup{justification=centering}
\caption{The Details of Three Datasets.}
\label{dataset}
\vspace{-0.2cm}
\input{document/table/dataset}
\vspace{-0.4cm}
\end{table}
\vspace{-0.3cm}
\subsubsection{Baseline Algorithms}
\vspace{-0.1cm}
We compared our model with seven Text Classification (TC) methods including TextCNN~\cite{textcnn}, DPCNN~\cite{dpcnn},  FastText~\cite{fasttext1,fasttext2},  TextRNN~\cite{textrnn},  TextRNN-Attn~\cite{textrnnattn},  TextRCNN~\cite{textrcnn},  Trasnformer~\cite{vaswani2017attention} and three state-of-the-art HMC approaches including  HMCN-F,  HMCN-R~\cite{wehrmann2018hierarchical}, and  HARNN~\cite{huang2019hierarchical}. We choosed three Hierarchical Multi-label Text Classification (HMTC) methods including HE-AGCRCNN\footnote{\url{https://github.com/RingBDStack/HE-AGCRCNN}}\cite{peng2019hierarchical}, HiLAP\footnote{\url{https://github.com/morningmoni/HiLAP}}\cite{mao2019hierarchical}, two variants of HiAGM\footnote{\url{https://github.com/Alibaba-NLP/HiAGM}}\cite{zhou2020hierarchy} (i.e., HiAGM-LA and HiAGM-TP).
We also posted three ablation models of HIRPCN as baselines, including \textit{w/o Interdisciplinary Graph} (w/o IKE),  \textit{w/o Hierarchical Transformers} (w/o SIE), and  \textit{w/o ALL}. 
We summarized four critical attributes of these baseline models in Table~\ref{overall} as:
\begin{itemize}
    \item \textbf{Hierarchical Awareness:} represents whether the model uses the hierarchical information of the label structure to enhance the model performance. 
    \item \textbf{Text-type Awareness:} represents whether the model is aware of the document type within each part of the research proposal. 
    \item \textbf{Interdisciplinary Awareness:} represents whether the model can utilize the interdisciplinary knowledge in the discipline system, i.e., the interdisciplinary graph. 
    \item \textbf{Start Prediction From Anywhere:} represents whether the model can start prediction with a given partial label, which is critical when aiding applicants and administrators in judging the proposal belongings.
\end{itemize}

\vspace{-0.4cm}
\subsubsection{Evaluation Metrics}
\vspace{-0.1cm}
To fairly evaluate the HIRPCN with other baselines, we adopted several metrics from the Multi-label Classification~\cite{metric1,metric2,xiao2021expert}, i.e., the \textit{Micro-F1} (MiF1), and \textit{Macro-F1} (MaF1). We conducted a 5-fold cross-validation and reported the average recommendation performance. 
\vspace{-0.3cm}
\subsubsection{Hyperparameters, Source Code and Reproducibility}
We set the SIE layer number $N_e$ to 8, the dimension size $h$ to 64, and the multi-head number to 8. The IKE layer number $N_g$ is set to 1. The IF layer number $N_d$ is set to 8, and the multi-head number is set to 8. We pre-trained the word embedding with a dimension ($h$) 64 via Word2Vec\cite{mikolov2013distributed} model. For the detail of training, we use Adam optimizer\cite{kingma2014adam} with a learning rate of $1\times 10^{-3}$, and set the mini-batch as 512, adam weight decay as $1 \times 10^{-7}$. The dropout rate is set to $0.2$ to prevent overfitting. The warm-up step is set as 1000. The $\alpha$ and $\beta$ for interdisciplinary graph construction set to 0.1.
We shared the source code via Dropbox\footnote{\url{https://www.dropbox.com/sh/x5m1jlcax8jp6tk/AAAg-KrGM8cHZuqoSfHKLRC0a}}.
\vspace{-0.3cm}
\subsubsection{Environmental Settings}
All methods are implemented by PyTorch 1.8.1\cite{paszke2019pytorch}. The experiments are conducted on a CentOS 7.1 server with a AMD EPYC 7742 CPU and 8 NVIDIA A100 GPUs. 
\vspace{-0.35cm}
\subsection{Experimental Results}
\begin{table*}
\centering
\captionsetup{justification=centering}
\caption{Experimental results on all dataset. The best results are highlighted in \textbf{bold}. The second-best results are highlighted in \underline{underline}.}
\vspace{-0.2cm}
\label{overall}

\input{document/table/overall_withnotation}
\vspace{-0.4cm}
\end{table*}
\subsubsection{\textbf{RQ1:} Overall Comparison}
% This experiment aims to answer: \textit{How is the performance of HIRPCN compared to each baseline method?}
We evaluated every method on all datasets and reported the results in Table~\ref{overall}. 
To compare the overall performance of the different models, we organized each step's prediction results flatly. 
Then, we used the \textit{MiF1} and the \textit{MaF1} as the overall evaluation metrics. From the results, we can observe that:
Firstly, the HIRPCN achieved the best performance among all datasets, proving that HIRPCN could better solve the challenges mentioned in the IRPC task. 
Besides this, the methods with top-down HMC schema (e.g., HMCN, HARNN, HIRPCN, and its variants) universally performed better than TC methods. 
The first reason is that the methods with top-down HMC schema organize the labels in a hierarchical view rather than a flat view, which can reduce the candidate label number in each step. 
Secondly, one of the critical attributes of the top-down HMC schema is that the model will generate the current prediction base on its previous prediction state. Because there are fewer categories in the earlier levels, the classification in these levels will be easier to predict correctly. Thus, the precise former results will facilitate the performance of these methods in the rest steps. 
We can also observe that the HMTC methods (e.g., HE-AGCRNN, HiLAP, and HiAGM) performed better than the TC method but slightly worse than the HMC methods. The reason for the former is although their encoder is similar to the TC methods, the HMTC methods either adopted the hierarchical structure into loss function (e.g., the HE-AGCRNN), initialized the label embedding by hierarchical structure (e.g., the HiAGM and its variants), or explicitly using the dependency of each label(e.g., the HiLAP). The latter is because we adopted the Hierarchical Transformer-based encoder for semantic information extraction in these HMC methods, which is more suitable for modeling long text.
Further, we noticed that the model's performance on the three datasets gradually declined. 
As we mentioned before, the interdisciplinary proposals have more labels on each level, making the classification task harder.
The performance on RP-bi is better than that on RP-differ, indicating due to the disparity between the two major disciplines, identifying the topic path under them will become much more challenging. Also, the performance of HIRPCN is better than other baselines, showing its robustness and generalization on the datasets with different interdisciplinarity.

\vspace{-0.25cm}
\subsubsection{\textbf{RQ2:} Ablation Study}
\vspace{-0.05cm}
% This experiment aims to answer: \textit{Is each component of HIRPCN valid?}
We also discussed the performance of each ablation variant of HIRPCN. 
The w/o SIE replaced the SIE component with a vanilla Transformer, making the model unable to discriminate the different types of documents (no \textbf{Text-type Awareness}).
The w/o IKE removed the IKE component. In each prediction step, w/o IKE will initialize the input previous prediction results by a look-up table  (no \textbf{Interdisciplinary Awareness}). 
The w/o ALL ablated both the IKE and SIE to represent the 
blank control group. 
From the Figure~\ref{overall}, we can observe that:
First of all, the HIRPCN beats all its ablation variants. This observation proves that both SIE and IKE component is valid in our model design. We also noticed that w/o SIE performs worse than w/o IKE in each dataset (e.g., from -3.8\% to -8.2\% deterioration on Micro-F1). This phenomenon indicates that the \textbf{type-specific semantic information} will play a more critical role in topic path detection. 
Secondly, w/o IKE performed slightly worse (e.g., -1.5\% deterioration on Micro-F1) than HIRPCN on the RP-all dataset, while this deterioration (e.g., -1.9\% and -9.1\% in Micro-F1) becomes larger on RP-bi and RP-differ. This phenomenon proves that the \textbf{interdisciplinary knowledge} plays a significant role in the classification of interdisciplinary proposals.
Lastly, HIRPCN and w/o IKE, both with hierarchical Transformer, are significantly superior to other HMC methods on RP-all and RP-bi, showing the advantage of modeling \textbf{various lengths of the documents} and the \textbf{multiple types of text} for proposal embeddings. 

\vspace{-0.25cm}
\subsubsection{\textbf{RQ3:} Level-wise Performance}
\vspace{-0.05cm}
\begin{table}
\centering
\captionsetup{justification=centering}
\caption{Experimental results on RP-differ in level-wise. The best results are highlighted in \textbf{bold}. The second-best results are highlighted in \underline{underline}.}
\label{RPdiffer}
\vspace{-0.20cm}

\input{document/table/differ-f1}
\vspace{-0.6cm}
\end{table}

\begin{figure*}[!h]
    \centering
    \includegraphics[width=0.95\linewidth]{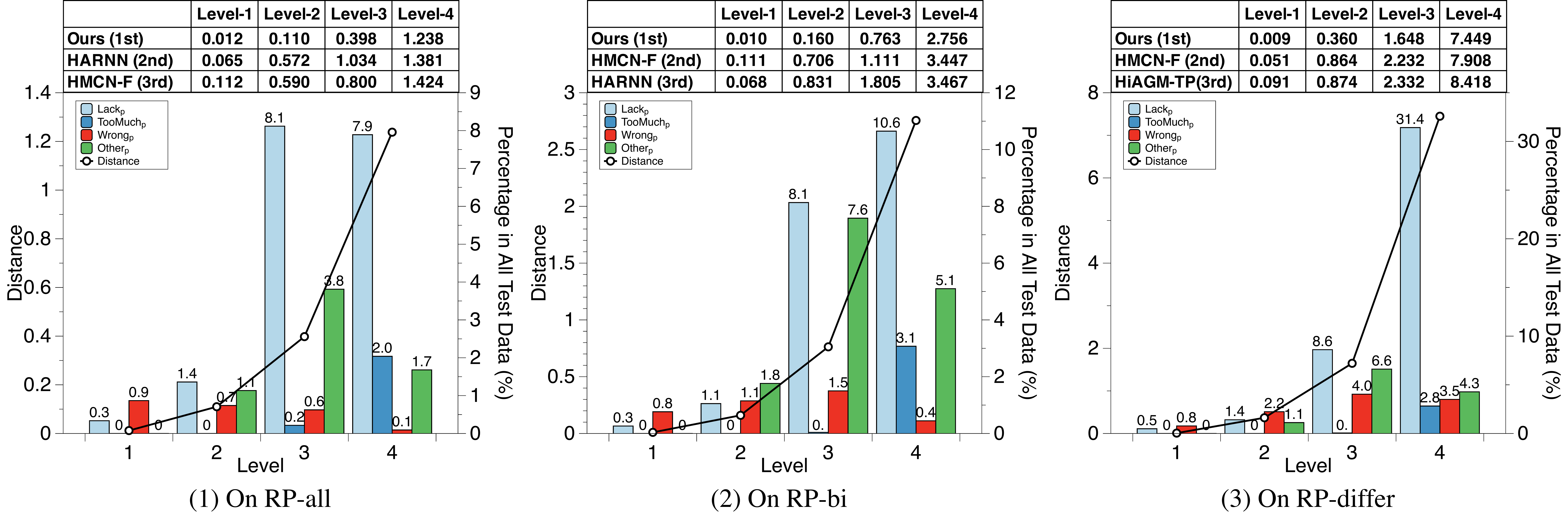}
    \vspace{-0.2cm}
    \captionsetup{justification=centering}
    \caption{The details of the wrong case and level-wise prediction distance on three datasets. The attached table also reported the interdisciplinary distances of the second and third-best methods.}
    \vspace{-0.4cm}
    \label{fig:stat_wrong}
\end{figure*}
% \note{to discuss why the level results can reflect the granularity.}
%We further evaluate the ability of HIRPCN and other baselines to give the proper granularity of prediction. We use the level-wise MiF1 and level-wise MaF1 as metrics to indicate the performance on granularity of the prediction. We compare HIRPCN with all baselines on RP-differ. Figure~\ref{RPdiffer} shows that HIRPCN outperforms other baseline methods and variants on each level. We also observe that the accuracy of each model tends to decrease, which is caused by the number of categories increasing rapidly when the hierarchical level depth increases. The results on every level prove that HIRPCN could predict the best \textbf{granular} on IRPC task.
% 1/ level 结果都好
% 2/ 探讨下降原因HMC方法
% 3/ 下降幅度更低

%我们进一步展示了模型和baseline细粒度的分类结果 by report他们每一层的分类结果。图2展示了结果。我们发现每个level结果都好，证明我们的模型
% This experiment aims to answer: \textit{How is model performance on each level?}
We reported the performance of level-wise prediction of HIRPCN and other baselines on the RP-differ dataset. 
Table~\ref{RPdiffer} showed the level-wise Micro-F1 and level-wise Macro-F1 classification results on four levels.
From the results, we can observe that  HIRPCN outperforms all the baseline methods and the variants on every level, showing the advanced ability of the classification at all granularity.
We also observed that the performance of each model tends to decrease with the depth of level increase. The reason is that the number of categories increases rapidly, making the classification task harder.
% Lastly, w/o IKE demonstrates a relatively slow decreasing tendency compared with other variants, proving the advantage of incorporating interdisciplinary knowledge for fine-grained discipline prediction. 
We also noticed that Macro-F1 dropped relatively more suddenly than Micro-F1 on each method. From the dataset statistic, we interpreted this as: 1) the length of the label path is variable, so in some cases, the last two levels might contain no target. 2) the test set is imbalanced, so some disciplines might not contain samples. Thus, the Macro-F1 will be lower than Micro-F1, especially in the last two levels.
The level-wise model performance on RP-all and RP-bi show the same tendency as on RP-differ so we won't go into detail.

\vspace{-0.3cm}
\subsubsection{\textbf{RQ4:} Evaluation of Prediction Results}
% This experiment aims to answer: \textit{How good is the prediction result?} 
The Micro-F1 and the Macro-F1 can measure how accurate the prediction is. However, it might be insufficient to evaluate the HIRPCN in real-world applications. In this experiment, we conducted two statistical analyses on three datasets for prediction results. The first analysis evaluates how close the prediction is to the truth label. For instance, if given a truth label in the 3rd level as $\{\textit{A0101}\}$, a prediction $\{\textit{A0102}\}$ might be more acceptable than $\{\textit{B0101}\}$, although they are both wrong.  Based on that, we developed a metric named interdisciplinary distance given as:
\begin{equation}
    % \mathcal{D}_{i}() = \sum_{l \in L_i}min(\{{d(l, \hat{\mathbb{L}}\})}),
    \mathcal{D}(L_k,\hat{L}_k) = \sum_{l \in L_k}min(\left\{dis(l, \hat{l})\right\}_{\hat{l}\in\hat{L}_k}),
\end{equation}
where $L_k$ and $\hat{L}_k$ are the truth labels and the prediction results in level-$k$, respectively. 
$min(\cdot)$ will take the minimized value from a given set. 
We adopted a revised edit distance as the distance function $dis(\cdot)$. 
In detail, if any two-digit or major discipline in the predicted ApplyID code is different from the truth, the function will instantly return the penalty. 
By discussing it with funding agencies, we set the penalty as $\{1, 10, 30, 50\}$, depending on the remaining step. 
For example, the distance between \textit{A0101} and \textit{A0102} will be $1$ due to they are different in level-$3$, and the remaining step will be $0$. 
And the distance between \textit{A0101} and \textit{B0101} will be $30$ because they are already different in the first level and their remaining step will be $2$. 
Also, if $l$ or $\hat{l}$ is null (i.e., the $L_k$ or $\hat{L}_k$ is empty), the distance will be $k$. 
Thus, $\mathcal{D}(L_k,\hat{L}_k)$ will quantitatively assess the distance between truth label set $L_k$ and predicted set $\hat{L}_k$. Then, we level-wise average the distance of every sample from the test set. The higher  average interdisciplinary distance represents the prediction for this level-$k$ are more offset.
Another analysis is we count the appearance number of three wrong cases, i.e., \textbf{Lack}, \textbf{TooMuch}, \textbf{Wrong}, and \textbf{Other}.
\textbf{Lack} represents the prediction in the current level is null, which means the model stop at a coarse level.
\textbf{TooMuch} means the model step into a too-fine-grained level, and the truth label in the current level is null.
\textbf{Wrong} represents that the model gives a prediction that not following the hierarchy. 
\textbf{Other} represents all other wrong cases. 

We calculated the average distance of each level and counted the number of wrong cases. The results of these analyses were reported in Figure~\ref{fig:stat_wrong}. 
Firstly, we can observe that the interdisciplinary distance and the number of wrong cases grow from level-1 to level-4, which is the same as the observation of the decline of Micro-F1 and Macro-F1 in Table~\ref{overall}. 
% Secondly, we noticed that the maximum distance is in the 4th level of RP-differ, with $7.449$, which means approximately only the last two digits are wrong on average (e.g., truth label in level-4 is \textit{A010101}, and model predict \textit{A010102}). 
Secondly, HIRPCN outperforms the second and third-best methods in the distance and shows excellent superiority in the first few levels. This observation shows the same phenomenon as the comparison with Micro-F1 and Macro-F1 in Table~\ref{overall}. 
% This phenomenon shows that although the prediction in the last few levels is hard, the model can give acceptable results. 
Furthermore, due to the most happening wrong cases in the level-4 of RP-differ is \textbf{Lack}, this distance is caused by the model stopping prediction prematurely rather than giving an incorrect prediction, which is acceptable in the real world. 
Apart from those, we noticed that the most reason to cause the distance between predictions and truth labels is \textbf{Lack}.  Simultaneously, the appearance number of \textbf{TooMuch} is relatively less, showing that our model will be more prude when predicting the topic path. Additionally, \textbf{Wrong}  happened around 0.7\% to 4.0\% in percentage in each level, showing that although we did not provide an explicit coherence constraint, the HIRPCN will instinctively learn the dependency and hierarchy between each level.

% \subsection{Case Study}
% We want to thoroughly analyze the model and discuss the potential for applying an online system. During the case study, we introduce three unique model features which can significantly improve the ideology of the funding application: 1) Hidden Interdiscipline Find 2) Prediction with Expert Knowledge 3) The Granularity. Suppose we are a newbie to apply for NSFC funding, and we might have no idea about the whole complex discipline system or the ApplyID code. The HIRPCN will provide the recommended discipline and help scientists choose the ApplyID code level by level with proper granularity. On the funding administrator side, they hope the HIRPCN could automatically mark the interdisciplinary proposals and the related disciplines to further decide on reviewer assignment.
\vspace{-0.3cm}
\subsubsection{\textbf{RQ5:} Prediction with Given Labels}
\begin{figure}
\vspace{-0.3cm}
\centering
\captionsetup{justification=centering}
\includegraphics[width=0.48\textwidth]{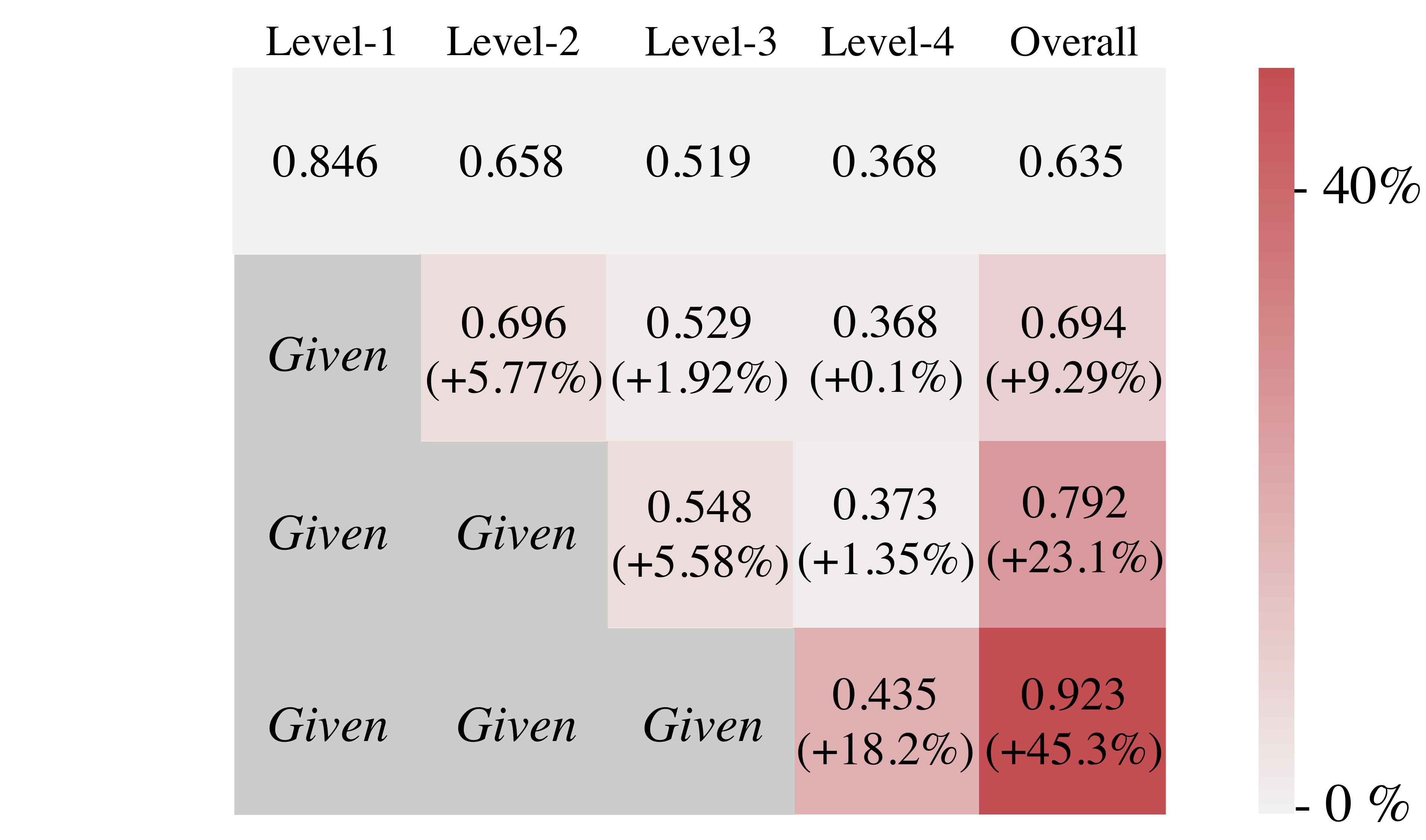}
\vspace{-0.2cm}
\caption{The performance and improvement with given labels in terms of Micro-F1.}
\label{Fig.partial_label}
\vspace{-0.7cm}
\end{figure}
\begin{figure*}[!h]
% \vspace{-0.4cm}
\centering
\captionsetup{justification=centering}
\includegraphics[width=0.95\textwidth]{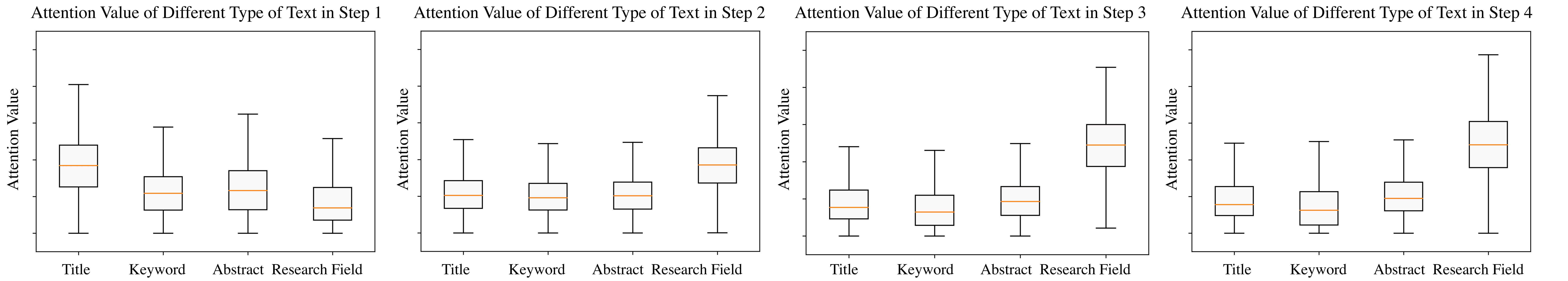} 
\vspace{-0.2cm}
\caption{Attention value of different type of semantic information in each step.}
\label{Fig.attn_type}
\vspace{-0.4cm}
\end{figure*}

\begin{figure*}[!h]
% \vspace{-0.1cm}
\centering
\captionsetup{justification=centering}
\includegraphics[width=0.95\textwidth]{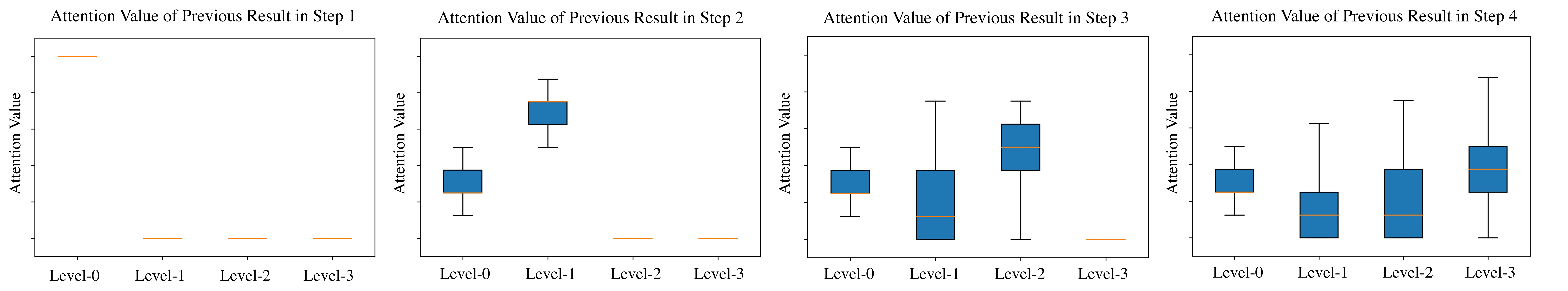} 
\vspace{-0.2cm}
\caption{Attention value of historical prediction results in each step.}
\label{Fig.attn_pre}
\vspace{-0.6cm}
\end{figure*}
% This experiment aims to answer: \textit{How will the Start From Anywhere characteristics improve the model performance by providing partial labels from applicants?}
The \textit{Given Labels} is a partial label set provided by applicants or funding administrators when they try to fill the ApplyID. 
As mentioned in the previous section, we hope the HIRPCN can be an assistant who knows the whole picture of the discipline system to aid the categorization with the manager. 
This study evaluated the improvement when HIRPCN receives incomplete labels and predicts the rest label. 
It is also worth noting that other baseline methods use a hidden vector to represent previous prediction results, which cannot initialize the prediction from a given partial label set. 
Thus, we compared the performance of HIRPCN with different levels of provided expert knowledge on the RP-differ dataset. 
As you can see in Figure~\ref{Fig.partial_label}, the first row of this heatmap showed the Micro-F1 of the HIRPCN on RP-differ when applicants were not given the partial labels. 
The rest rows illustrated when the first level, first and second level, first, second and third level were given and how the model improved its performance using these partial labels. 
The color of each cell showed the strength of the improvement. From the heatmap, we can observe that:
Firstly, with the incomplete labels given, the overall performance of HIRPCN rises. This phenomenon proves that the architecture enables the model to utilize the partial label to improve the overall performance.
Secondly, we noticed that the given partial labels could significantly improve the next-level prediction but decline in the remaining level. We interpret this phenomenon as the current prediction result might be more related to its most recent-level results. So, the model would be trained to raise a high attention on the most current level rather than other earlier-level. In summary, the improvement from the given partial label shows the potential application scenario for the AI assistant system.

\vspace{-0.2cm}
\subsubsection{\textbf{RQ6:} Attention Mechanism On Information Fusion} 
\vspace{-0.1cm}
% This experiment aims to answer: \textit{Can HIRPCN adaptively utilize the semantic information and historical prediction results?}
Adaptively utilizing  each part of the research proposal and the previous prediction results is the most critical idea of the \textit{Information Fusion} component. 
In this experiment, we gathered the model's attention value on each previous prediction step and each different type of text from the IF component. 
The results was reported in Figure~\ref{Fig.attn_type} and Figure~\ref{Fig.attn_pre}. 
% \begin{itemize}

\noindent \textbf{The attention to different types of text:}
Figure~\ref{Fig.attn_type} illustrated the attention value of the different types of documents in each step. 
We observed that the \textit{Title} field is important while predicting the major discipline (step 1). However, HIRPCN will pay more attention to the \textit{Research Field} and \textit{Abstract} to generate the rest labels. 
Although, it might not be noteworthy to interpret and compare the model mechanism with experts' behavior. But, compared to simply concatenating the text sequence, the attention mechanism could provide more potential for information fusion.

\noindent \textbf{The attention to historical prediction results:}
Figure~\ref{Fig.attn_pre} showed the model's attention to the historical prediction results. We noticed that the model would adopt the interdisciplinary knowledge information from the most recent predictions. However, the model might consider other historical predictions to generate the rest in some cases. This phenomenon also proves the experiment observation in \textbf{RQ5} (i.e., the given partial label will significantly improve the next level's performance).

% \end{itemize}
% \noindent

% \noindent
To sum up, the attention in the Information Fusion will adaptively variate during the prediction progress. This phenomenon shows the validity of the model component and brings excellent potential to the HIRPCN.

\begin{figure}[!t]
\centering
\captionsetup{justification=centering}
\includegraphics[width=0.48\textwidth]{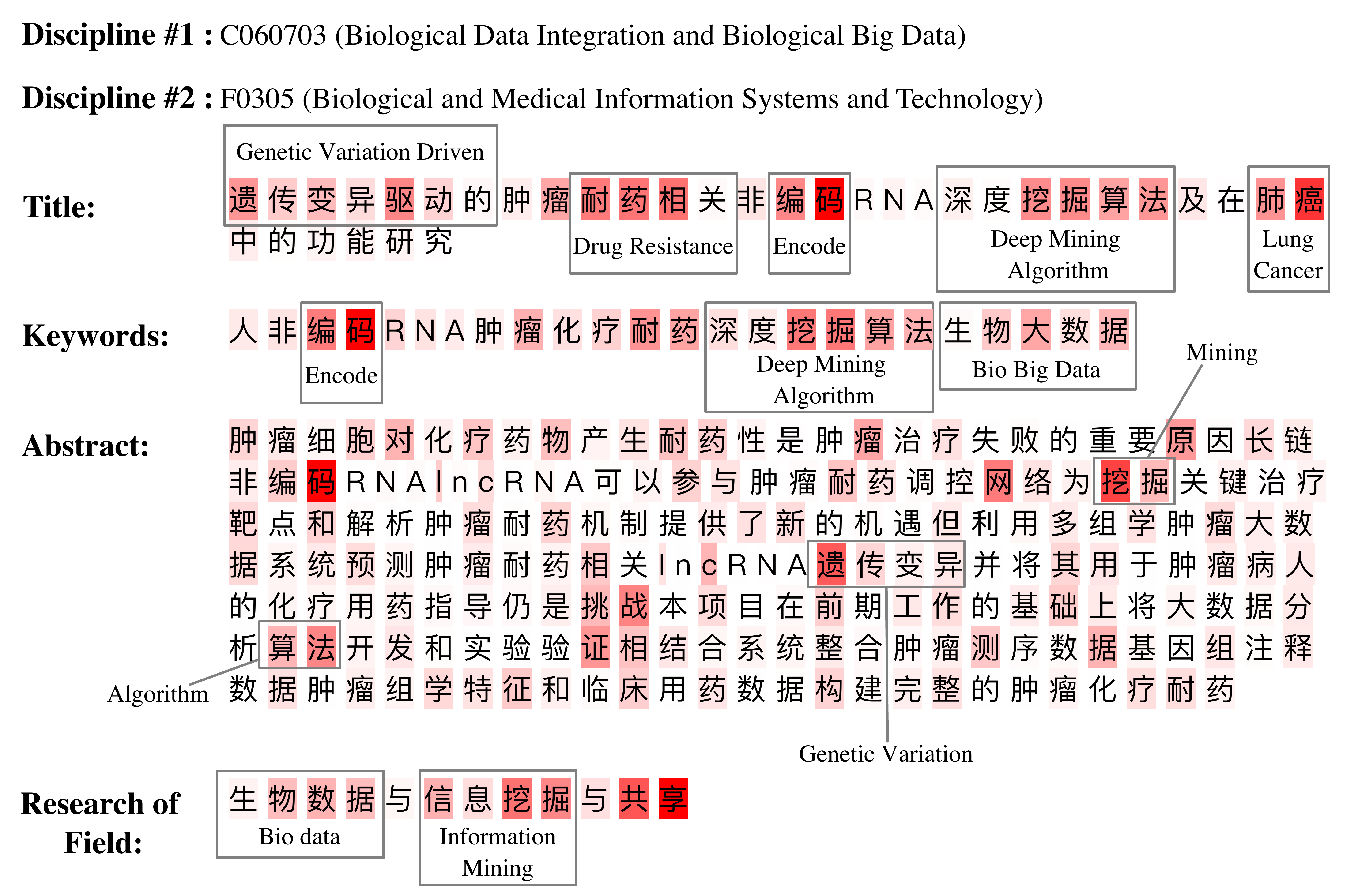} 
\vspace{-0.2cm}
\caption{A interdisciplinary research proposal and its word-level attention visualization from the \textit{Information Sciences} and the \textit{Life Sciences}.}
\vspace{-0.6cm}
\label{Fig.attn_text}
\end{figure}
\vspace{-0.3cm}
\subsubsection{\textbf{RQ7:} Attention Mechanism On Text}
\vspace{-0.15cm}
% This experiment aims to answer: \textit{How could the attention mechanism on text enhance the model performance?}
Figure~\ref{Fig.attn_text} is a case study of an interdisciplinary research proposal. If a word token gained more attention from SIE, it would be shaded redder in this figure. 
The first observation is that the SIE raised a high signal on the critical words instead of the stop words, even in the long text data. For example, in the \textit{Abstract}, the term “挖掘” (Mining) and “算法” (Algorithm) generally gained more attention than the word “是” (Is) and “和” (And). We think the attention mechanism will help the HIRPCN learn the importance of each word token and find the critical term to identify the topic path. 
Further, the attention value was high for the terms relevant to the \textbf{corresponded discipline} of the research proposal. For example, in the \textit{Title}, the word “肺癌” (Lung Cancer) and “深度挖掘算法” (Deep Mining Algorithm) gained more attention due to they were relevant respectively to the \textit{Life Sciences} and \textit{Information Sciences}.
This phenomenon shows that the HIRPCN can extract useful semantic information from documents by highlighting domain-related words.  
Lastly, the self-attention mechanism worked well in the documents with multiple types and variable lengths, such as the term “耐药相关” (Drug Resistance) in the \textit{Title} or the words “挖掘” (Mining), “算法” (Algorithm) in the \textit{Abstract}. 
We believe this phenomenon shows why the SIE is superior for handling the \textbf{multiple texts} in research proposals. In addition, those highlighted critical words can bring interpretability to HIRPCN, which is also essential for fund administrators. 

% \vspace{-0.3cm}
\begin{figure}[!t]
\vspace{-0.3cm}
\centering
\captionsetup{justification=centering}
\includegraphics[width=0.45\textwidth]{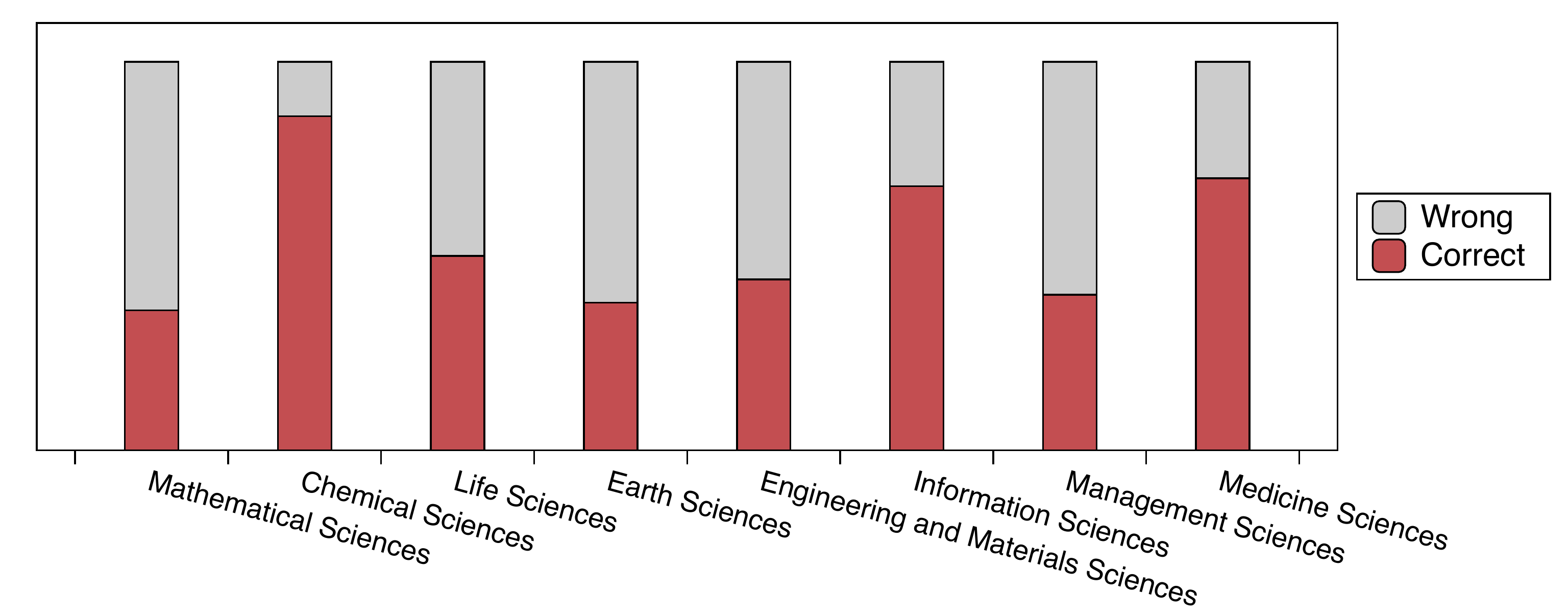}
\caption{The results of hidden interdiscipline find study.}
\label{Fig.evalresult}
\vspace{-0.5cm}
\end{figure}
\vspace{-0.3cm}
\subsubsection{\textbf{RQ8:} Hidden Interdiscipline Find} 
\vspace{-0.1cm}
% This experiment aims to answer: \textit{What is the supplementary generated label significance to the peer-review assignment?}
Among all the wrong cases in RP-all, the HIRPCN classifies part of non-interdisciplinary research to interdisciplinary, which means except for their \textbf{labeled discipline}, our model predicts an \textbf{extra discipline}. 
After observing those wrong cases, we noticed that these non-interdisciplinary research proposals were somewhat related to their extra predicted disciplines. 
To understand this phenomenon, We invited eight fund administrators from NSFC with an insightful understanding and management experience of the eight major disciplines to judge if those extra disciplines can provide helpful information for improving the reviewer assignment.
% 评价这些额外预测的学科信息能否用于更公平的审稿人分配
We first group those cases by the extra discipline predicted by HIRPCN. Then we sample 50 research proposals from each group. We send those research proposals to each expert according to the expert's familiar discipline. 
% 专家会标注这个额外的学科是否有用 yes no
If the expert thinks the reviewers from the extra discipline should be considered to supplement the peer-review group, they will mark it as \textit{correct}, otherwise \textit{wrong}. We illustrated the percentage of \textit{correct} of each extra discipline in Figure~\ref{Fig.evalresult}. From the results, we can observe that those experts marked 57\% of the samples as \textit{correct}. Further, all experts commented that the HIRPCN could detect the hidden interdisciplinary research domain. 
Besides these, some experts commented that although some cases were marked \textit{wrong}, most of them explicitly use the keywords from the predicted extra discipline. The Experts categorized those cases into non-interdisciplinary research because their focal point is on the labeled discipline instead of the predicted extra one (e.g., a study on \textit{Artificial Intelligence} might introduce the ideas and terms from \textit{Neural Sciences}. Still, the neuroscientists might be not suitable to review it). 
Furthermore, some experts commented that they marked some cases as \textit{wrong} because the investigators created the labeled discipline to cover the extra discipline topic. So, the labeled discipline is enough to describe the research domain. For example, the discipline \textit{Bionics and Artificial Intelligent} (C1005) in \textit{Life Sciences} (C), was created to represented the interdiscipline of \textit{Bionics} (C10) and \textit{Artificial Intelligent} (F06). 

% (1) 

% % 有些专家评论，尽管有些case是错的，但他们在一定程度上展示了和predicted label的相关性，这些相关的内容不作为主要的研究内容，因此不是非常属于predicted label。这证明了我们的模型能够发现一些弱相关性。
% (2) 

% % (2) Some experts comment that although some parts of the \textit{wrong} samples explicitly use the keywords from the predicted extra discipline. However, from their perspective, they cannot categorize them into interdisciplinary research due to their methodology or idea having little relation with the predicted discipline.  
% % \note{revise the little}
% % 

% % 学科是进化的，有些学科是融合的。专家评论有些错例是因为，given的学科本身是交叉学科，且given的学科包含了predicted的学科，因此不需要给她指定这个predicted学科。例如，C1005 包含了A和B，而我们的模型预测出来B，
% (3) 

% (3) Some experts comment that they marked some cases as \textit{wrong} because the labeled discipline is more suitable to describe the research domain even if the given one and the predicted extra one were interdisciplinary in the past. The ApplyID is an evolving and dynamic discipline system. For example, there were no ApplyID codes represent the interdiscipline of \textit{Bionics} and \textit{Artificial Intelligent} in 2018. However, in 2020, there was a new discipline \textit{Bionics and Artificial Intelligent}, which is denoted by code \textit{C1005}, proposed to describe the related research field. 
% example -> 学科进化 讲清楚

\begin{figure}
\vspace{-0.1cm}
\centering
\captionsetup{justification=centering}
\includegraphics[width=0.48\textwidth]{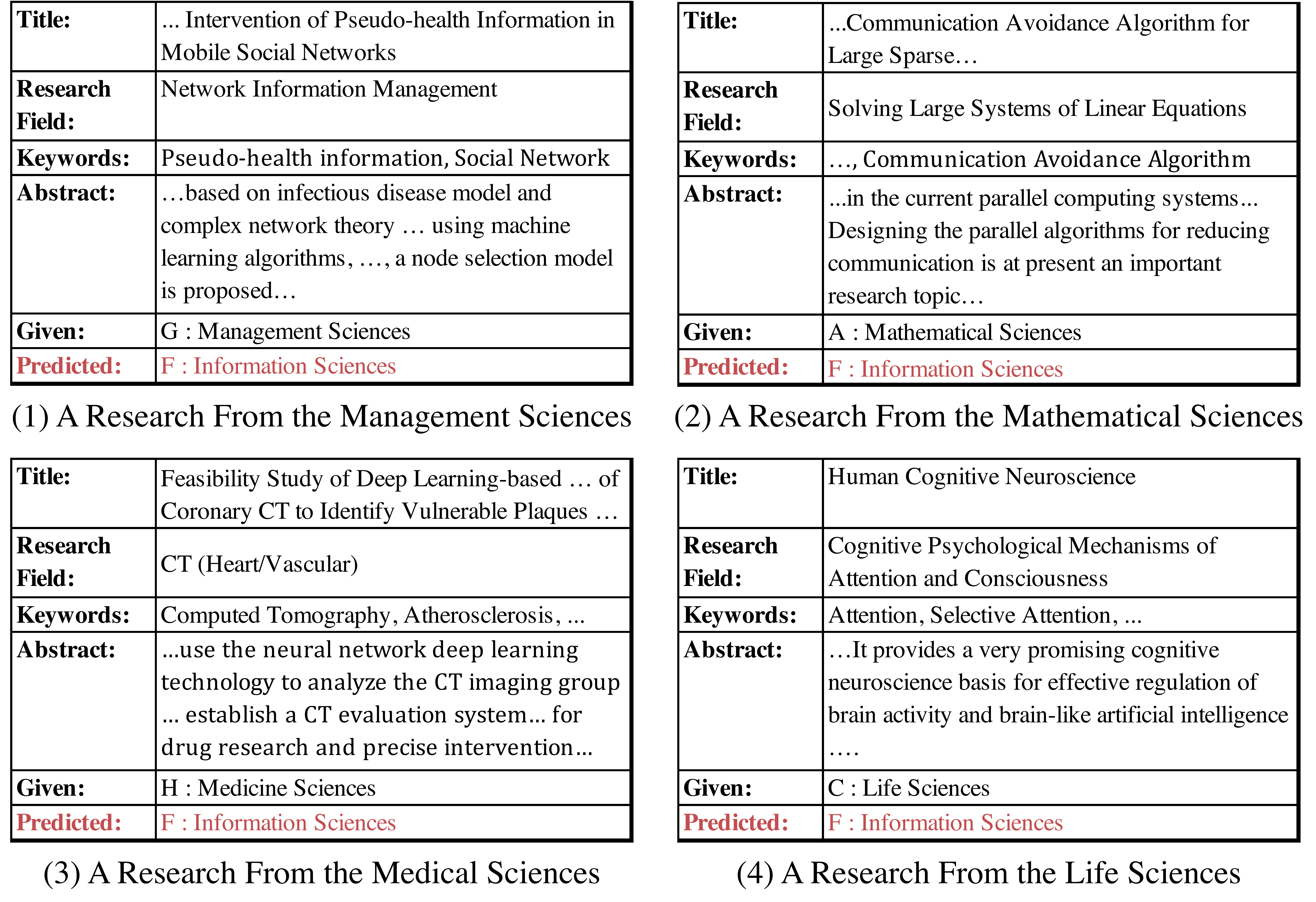}
\caption{Four cases in different area, while HIRPCN detect another hidden discipline.}
\label{Fig.wrong_case}
\vspace{-0.4cm}
\end{figure}

We randomly picked four proposals and listed them in Figure~\ref{Fig.wrong_case}. Applicants manually labeled those proposals with one discipline, but our model extra marked them with another domain from the Information Sciences. We can observe that those research proposals are highly related to their labeled disciplines, while they also strongly bond with the Information Sciences, whether the methodology or the idea. We believe that, with the HIRPCN, those research proposals can be reviewed by experts who are knowledgeable in both areas, thus providing a fair evaluation. 
\vspace{-0.3cm}
\subsection{Interdisciplinary Graph Settings}
\vspace{-0.1cm}
\begin{figure}[t]
\centering
\captionsetup{justification=centering}
\includegraphics[width=0.35\textwidth]{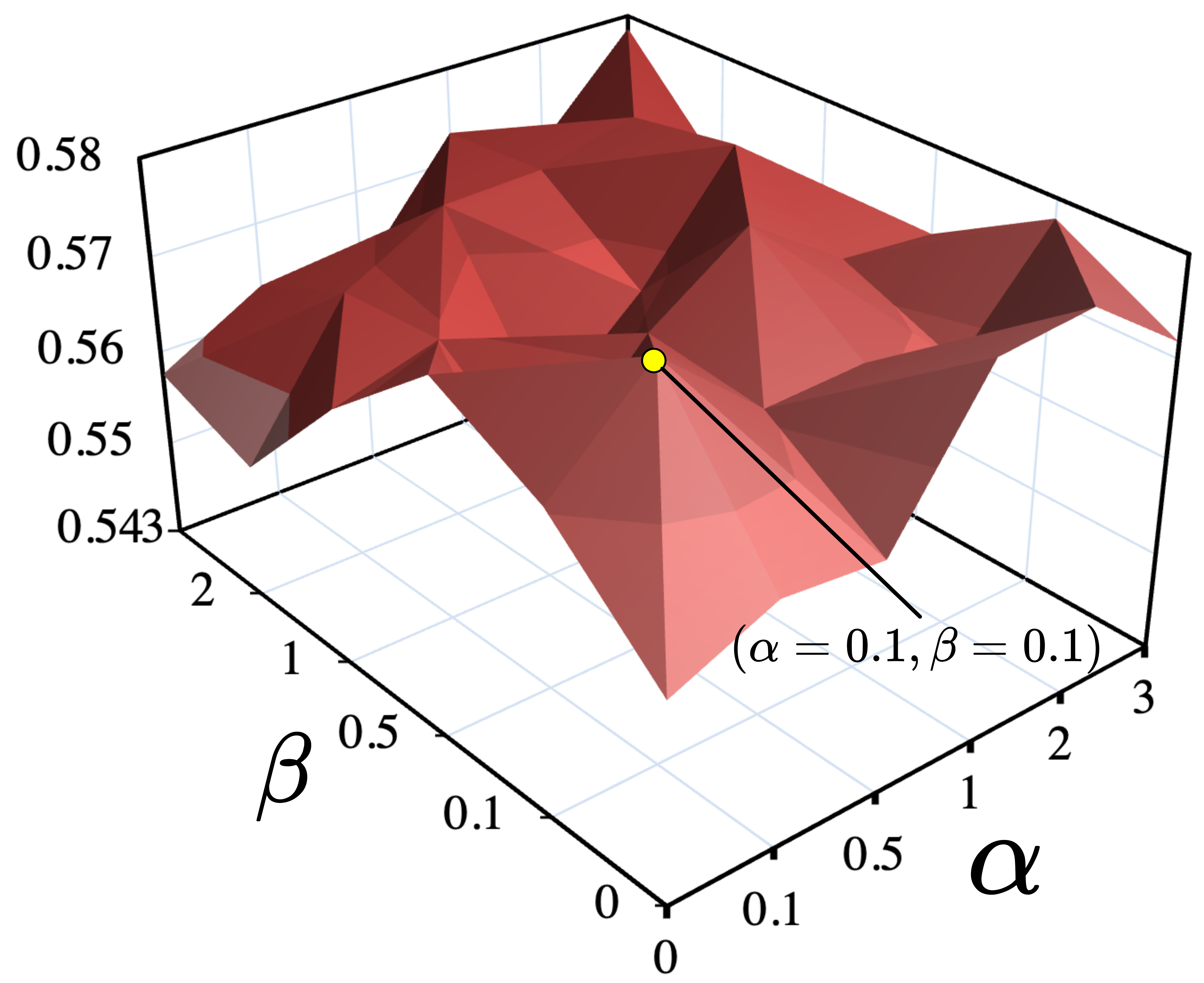} 
\caption{The model performance with different interdisciplinary graph settings.}
\label{Fig.alpha_beta}
\vspace{-0.4cm}
\end{figure}
As mentioned in Section~\ref{co}, we introduced a weighted graph to represent the interdisciplinarity between each discipline. This weight is defined by the proportion (\textbf{disparity}) and the citation number (\textbf{frequency}) of the overlap topic  in Equation~\ref{g_equation}.
On the one hand, if two domains show high interdisciplinarity, their research area will have great distinction overall, but their overlap topic will be a significant hot area to research.  
On the other hand, two disciplines that have too many overlapped topics or their overlap topic have little citation might be two similar disciplines (e.g., the \textit{Machine Learning} and the \textit{Deep Learning}) or two irrelative areas (e.g., the \textit{Management Sciences} and the \textit{Electronic Engineering}), both show little interdisciplinarity. We use two parameter $\alpha$ and $\beta$ to set the weight of two component. 

We trained the HIRPCN on RP-differ with different combinations of $\alpha$ and $\beta$. The result of model performance in epoch 100 is visualized in Figure~\ref{Fig.alpha_beta}. From the figure, we can observe that the best setting of the interdisciplinary graph is set the $\alpha$ and $\beta$ to 0.1.
If we both set the $\alpha$ and $\beta$ to 0, all edge weights will equal 1.0, and the graph will deteriorate to a co-topic graph. 
In epoch 100, this setting would undermine the model performance by around -3.5\% compared to the best setting. 
We believe this phenomenon will explain the importance of interdisciplinary knowledge and our adopted RS measurement.
\vspace{-0.45cm}
\subsection{Hyperparameter Selection}
\vspace{-0.1cm}
\begin{figure}[t]
\centering
\captionsetup{justification=centering}
\includegraphics[width=0.48\textwidth]{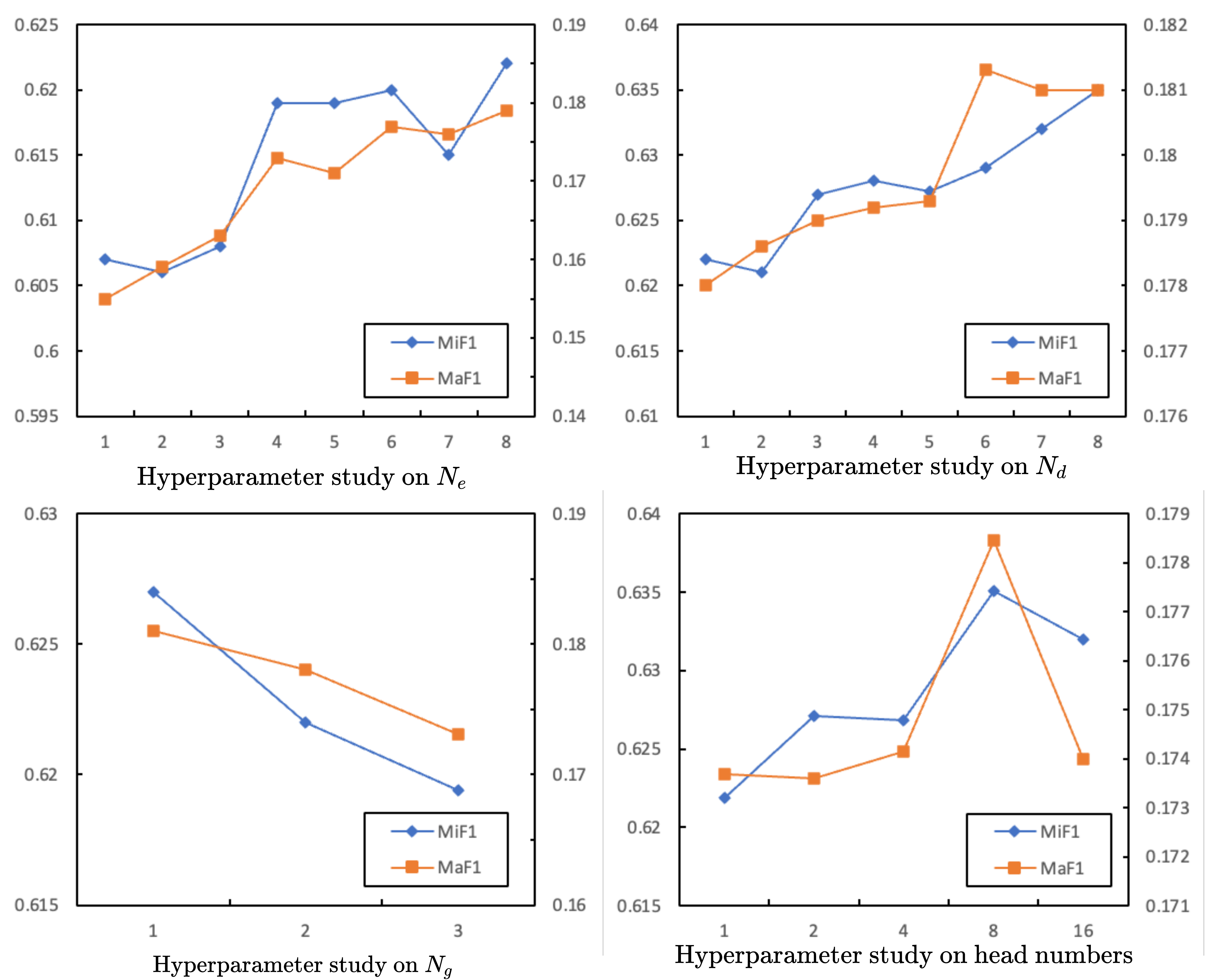} 
\caption{Hyper parameter selection study.}
\vspace{-0.6cm}
\label{Fig.hyper}
\end{figure}
We conducted experiments to evaluate the effect of four key hyperparameters, i.e., $N_e$, $N_d$, $N_g$, and the number of self-attention heads in the SIE and IF. We investigated the sensitivity of those parameters in each part of HIRPCN and reported the results in Figure \ref{Fig.hyper}. From the figure, we can observe that with the $N_e$ and $N_d$ vary from 1 to 8, the performance of  HIRPCN increases at first and then decreases slowly. The model will likely be over-fitting to impact the classification results with more parameters introduced.
Also, we found that the model would have the best performance when IKE sampled a one-hop neighborhood to extract interdisciplinary knowledge. The reason is that the interdisciplinary graph has many edges, and two or more hop sampling strategies will make the sampled sub-graph denser, which will provide unrelated knowledge for IKE.
Lastly, more attention heads will generally improve the performance of  HIRPCN. Meanwhile, we found that too many attention heads will make the model performs worse.
Based on the experimental result, we set $N_e$ to 8, $N_d$ to 8, $N_g$ to 1, and the number of self-attention heads to 8 to achieve the best performance. 

%% file: document/table/disciplines.tex
% \usepackage{graphicx}

\resizebox{0.95\linewidth}{!}{%
\begin{tabular}{cccccc}
\toprule
Prefix & Major Discipline Name              & Total & $|C_2|$ & $|C_3|$ & $|C_4|$ \\ \midrule
A      & Mathematical Sciences              & 318          & 6       & 57      & 255     \\
B      & Chemical Sciences                  & 392          & 8       & 59      & 325     \\
C      & Life Sciences                      & 801          & 21      & 162     & 618     \\
D      & Earth Sciences                     & 166          & 7       & 94      & 65      \\
E      & Engineering and Materials Sciences & 138          & 13      & 118     & 7       \\
F      & Information Sciences               & 100          & 7       & 88      & 5       \\
G      & Management Sciences                & 107          & 4       & 57      & 46      \\
H      & Medicine Sciences                  & 456          & 29      & 427     & 0       \\ 
- & Total Disciplines & 2478 & 95 & 1062 & 1321
\\\bottomrule
\end{tabular}%
}

%% file: document/table/dataset.tex
% \begin{tabular}{cccc}
% \toprule
% Dataset       &  RP-2020 \\ \midrule
% \multicolumn{1}{c}{Total Proposal Number}                   & 20632     \\
% \multicolumn{1}{c}{Avg. Labels Length}  & 3.397
%   \\
% \multicolumn{1}{c}{Avg. Labels Num in Level-1} & 2.000     \\
% \multicolumn{1}{c}{Avg. Labels Num in Level-2}  & 2.000     \\
% \multicolumn{1}{c}{Avg. Labels Num in Level-3}  & 1.985     \\
% \multicolumn{1}{c}{Avg. Labels Num in Level-4} & 0.808     \\
% \bottomrule
% \end{tabular}%
\resizebox{0.95\linewidth}{!}{%
\begin{tabular}{cccc}
\toprule
Dataset                                             & RP-all & RP-bi  & RP-differ \\ \midrule
\multicolumn{1}{l}{Total Proposal Number}                   & 280683 & 125466 & 20632     \\
\multicolumn{1}{l}{\#Avg. Labels Length} & 2.393  & 3.355  & 3.397
  \\
\multicolumn{1}{l}{\#Avg. Labels Num in Level-1} & 1.073  & 1.164  & 2.000     \\
\multicolumn{1}{l}{\#Avg. Labels Num in Level-2} & 1.197  & 1.442  & 2.000     \\
\multicolumn{1}{l}{\#Avg. Labels Num in Level-3} & 1.364  & 1.870  & 1.985     \\
\multicolumn{1}{l}{\#Avg. Labels Num in Level-4} & 0.477  & 0.726  & 0.808     \\
\bottomrule
\end{tabular}%
}

%% file: document/table/overall_withnotation.tex
\resizebox{0.95\textwidth}{!}{%
\begin{tabular}{lcccc|cccccc}
\toprule
\multicolumn{5}{l|}{Dataset}                                        & \multicolumn{2}{c}{RP-all}                                              & \multicolumn{2}{c}{RP-bi}                                               & \multicolumn{2}{c}{RP-differ}                                           \\ \midrule
Attributes               & \makecell{Hierarchical \\Awareness} &\makecell{Text-type\\ Awareness}& \makecell{Interdisciplinary\\ Awareness} &
\makecell{Start Prediction\\ From Anywhere} & MiF1          & MaF1 & MiF1            & MaF1 & MiF1           & MaF1 \\ \midrule
TextCNN               & \xmark                                                   & \xmark                                                  & \xmark                                                                     & \xmark                                                      & 0.456               & 0.168                         & 0.426               & 0.148                         & 0.438               & 0.063                        \\
DPCNN                   & \xmark                                                   & \xmark                                                  & \xmark                                                                     & \xmark                                                      & 0.376               & 0.108                         & 0.364               & 0.080                         & 0.346               & 0.020                        \\
FastText                & \xmark                                                   & \xmark                                                  & \xmark                                                                     & \xmark                                                      & 0.460               & 0.167                         & 0.427               & 0.149                         & 0.432               & 0.067                        \\

TextRNN                 & \xmark                                                   & \xmark                                                  & \xmark                                                                     & \xmark                                                      & 0.402               & 0.100                         & 0.364               & 0.077                         & 0.351               & 0.021                        \\

TextRNN-Attn & \xmark                                                   & \xmark                                                  & \xmark                                                                     & \xmark                                                      & 0.413               & 0.098                         & 0.367               & 0.072                         & 0.338               & 0.018                        \\
TextRCNN                & \xmark                                                   & \xmark                                                  & \xmark                                                                     & \xmark                                                      & 0.426               & 0.138                         & 0.381               & 0.100                         & 0.367               & 0.025                        \\
Transformer             & \xmark                                                   & \xmark                                                  & \xmark                                                                     & \xmark                                                      & 0.370               & 0.083                         & 0.339               & 0.066                         & 0.358               & 0.023                        \\ \midrule
HMCN-F         & \cmark                                                    & \xmark                                                  & \xmark                                                                     & \xmark                                                      & 0.676               & 0.459           & 0.582 & 0.308           & 0.569               & $\underline{0.161}$          \\
HMCN-R       & \cmark                                                    & \xmark                                                  & \xmark                                                                     & \xmark                                                      & 0.599               & 0.289                         & 0.506              & 0.185                         & 0.478               & 0.069                        \\
HARNN                   & \cmark                                                    & \xmark                                                  & \xmark                                                                     & \xmark                                                      & 0.686 & 0.443                         & 0.574               & 0.294                         & 0.516               & 0.102                        \\ \midrule

HE-AGCRCNN &\cmark &\xmark & \xmark & \xmark & 0.513               & 0.331                         & 0.467               & 0.289                         & 0.463               & 0.118                \\
HiLAP & \cmark & \xmark & \xmark & \xmark &  0.628               & 0.418           & 0.536 & 0.280           & 0.528               & 0.141 \\ 
HiAGM-LA& \cmark & \xmark & \xmark & \xmark & 0.594 & 0.400 & 0.505 & 0.258 & 0.501 & 0.134 \\
HiAGM-TP& \cmark & \xmark & \xmark & \xmark& 0.629 & 0.428 & 0.548 & 0.285 & 0.535 & 0.135\\
\midrule

w/o ALL      & \cmark                                                    & \xmark                                                   & \xmark                                                                     & \cmark                                  & 0.704            & 0.429                        
& 0.583             & 0.259 % epoch 270                         
& 0.512              & 0.098 % epoch 820 
\\
w/o SIE    & \cmark                                                    & \xmark                                                  & \cmark                                                                      & \cmark                                                       & 0.710               &  0.438                         & 0.587              &  0.276                        & 0.533 & 0.099   \\

w/o IKE      & \cmark                                                    & \cmark                                                   & \xmark                                                                     & \cmark                                                       &\underline{0.737}               & \underline{0.488}                         & \underline{0.636}               & \underline{0.329}                         & \underline{0.577} & 0.127                        \\
HIRPCN     & \cmark                                                    & \cmark                                                   & \cmark                                                                      & \cmark                                                       & \textbf{0.748}    & \textbf{0.512}              & \textbf{0.648}    & \textbf{0.384}              & \textbf{0.635}    & \textbf{0.180}             \\ \bottomrule
\end{tabular}}%

%% file: document/table/differ-f1.tex
\resizebox{0.48\textwidth}{!}{%
\begin{tabular}{lcccccccc}
\toprule
\multicolumn{1}{l}{Level} & \multicolumn{2}{c}{1}                                  & \multicolumn{2}{c}{2}                                  & \multicolumn{2}{c}{3}                                  & \multicolumn{2}{c}{4}                   \\ \midrule 
\multicolumn{1}{l}{Metric}                        & \multicolumn{1}{c}{MiF1} & \multicolumn{1}{c}{MaF1} & \multicolumn{1}{c}{MiF1} & \multicolumn{1}{c}{MaF1} & \multicolumn{1}{c}{MiF1} & \multicolumn{1}{c}{MaF1} & \multicolumn{1}{c}{MiF1} & \multicolumn{1}{c}{MaF1}\\ \midrule 
TextCNN                 & 0.735          & 0.600            & 0.403          & 0.285          & 0.254          & 0.067          & 0.190           & 0.040          \\
DPCNN                   & 0.689          & 0.543          & 0.284          & 0.151          & 0.132          & 0.017          & 0.084          & 0.009         \\
FastText                & 0.730           & 0.610           & 0.402          & 0.303          & 0.251          & 0.078          & 0.164          & 0.037         \\
TextRNN                 & 0.692          & 0.569          & 0.296          & 0.155          & 0.130           & 0.018          & 0.073          & 0.009         \\
TextRNN-Attn            & 0.677          & 0.548          & 0.278          & 0.148          & 0.125          & 0.015          & 0.071          & 0.008         \\
TextRCNN                & 0.690           & 0.559          & 0.310           & 0.183          & 0.165          & 0.025          & 0.105          & 0.010          \\
Transformer             & 0.702          & 0.577          & 0.308          & 0.183          & 0.135          & 0.022          & 0.070           & 0.009         \\ 
\midrule
HMCN-F                    & 0.782          & 0.670           & 0.565          & 0.461          & \underline{0.460}     & \underline{0.201}    & \underline{0.341}    & \underline{0.102}   \\
HMCN-R                  & 0.770           & 0.664          & 0.478          & 0.339          & 0.278          & 0.074          & 0.183          & 0.041         \\
HARNN                   & 0.793          & 0.683          & 0.529          & 0.409          & 0.341          & 0.122          & 0.226          & 0.058         \\ \midrule
HE-AGCRCNN                   & 0.743          & 0.693          & 0.439          & 0.411          & 0.289          & 0.132          & 0.171          & 0.068         \\ 
HiLAP & 0.732 & 0.624 & 0.523 & 0.418 & 0.419 & 0.172 & 0.306 & 0.082 \\
HiAGM-LA & 0.702 & 0.610 & 0.509 & 0.402 & 0.398 & 0.169 & 0.265 & 0.069 \\ 
HiAGM-TP & 0.731 & 0.627 & 0.538 & 0.430 & 0.429 & 0.178 & 0.314 & 0.084 \\
\midrule
w/o ALL                 & 0.810          & 0.692          & 0.564          & 0.441           & 0.357          & 0.133          & 0.219          & 0.060         \\
w/o SIE                   & 0.812 & 0.703 & 0.551 & 0.445 & 0.369 & 0.138 & 0.223 & 0.065\\
w/o IKE                 & \underline{0.825}    & \underline{0.715}    & \underline{0.627}    & \underline{0.506}    & 0.429          & 0.150           & 0.281          & 0.075         \\
HIRPCN                    & \textbf{0.846} & \textbf{0.731} & \textbf{0.658} & \textbf{0.540} & \textbf{0.519} & \textbf{0.222} & \textbf{0.368} & \textbf{0.114}\\
\bottomrule
\end{tabular}}%

%% file: document/related.tex
\vspace{-0.3cm}
\section{Related Works}
\subsection{Text Classification}
The text classification~\cite{li2020survey} problem is a classical natural language processing task, which aims to extract intrinsic semantic information and assign its related labels. 
The shallow approaches, such as Word2Vec~\cite{mikolov2013distributed}, and GloVe~\cite{pennington2014glove}, aim to map the word sequence into vector space and then classify each document by this representation. 
The deep-learning approaches developed many encoders to enhance the semantic information extraction, such as RNN-based~\cite{lstm, textrnn, textrnnattn}, CNN-based~\cite{textcnn, textrcnn, dpcnn}, Attention-based~\cite{vaswani2017attention}, and pretrained language model based~\cite{devlin2018bert}. 
Apart from those approaches, many solutions have been proposed to solve the Multi-label Text Classification, such as WRAP~\cite{yu2021multi}, LCBM~\cite{sun2020lcbm}, and 3WDLE~\cite{zhao2022selective}. 
However, in a real-world scenario, the labels of text classification are always exhibit a hierarchical structure, which named the Hierarchical Multi-label Text Classification (HMTC) task. To address this challenge, the study~\cite{peng2019hierarchical} learns the label dependency and uses this dependency to construct the loss function. HiLAP~\cite{mao2019hierarchical} proposed a reinforcement approach so the prediction can follow the label structure. HiAGM~\cite{zhou2020hierarchy} proposed two structure encoders for modeling the hierarchical labels. HiMatch~\cite{chen2021hierarchy} implemented a text-to-hierarchical-label matching network. 
The interdisciplinary research proposal classification task can be seen as a complicated form of HMTC task which aims to categorize multiple types of documents into a hierarchical discipline structure. 
\vspace{-0.3cm}
\subsection{Hierarchical Multi-label Classification}
Our work is most related to deep-learning based hierarchical multi-label classification~\cite{cerri2015hierarchical} (HMC)  methods which leverage standard neural network approaches for multi-label classification problems~\cite{giunchiglia2021multi,wehrmann2018hierarchical,gargiulo2019deep} and then exploit the hierarchy constraint~\cite{huang2019hierarchical} in order to produce coherent predictions and improve performance. 
Zhang et al.~\cite{Zhang2021} propose a document categorization method with hierarchical structure under weak supervision. The work~\cite{tang2020multi} design a attention-based graph convolution network to category the patent to IPC codes. The work in \cite{LaGrassa2021} use the convolution neural network as an encoder and explore the HMC problem in image classification. The works in \cite{nakano2020active} propose a active learning approach for HMC problem. Aly et al.~\cite{Aly2019} propose a capsule network based method for HMC problem. 
TAXOGAN~\cite{Yang2020} embedding the network nodes and hierarchical labels together, which focuses on taxonomy modeling. In recent studies~\cite{liu2021multi}, researchers try to exploit more information among the label embedding and label correlation.
Besides the label embedding approach, the study~\cite{giunchiglia2020coherent} explore the performance of hierarchy dependency as a training constraint and achieve well performance on a small dataset with neural networks.
The previous studies in this area are well constructed and have significant success in a specific dataset. However, few studies~\cite{xiao2021expert, xiao2022hmix} focus on the complex sci-tech text data classification with a hierarchical discipline structure.
In this paper, we inherit the fundamental idea of HMC networks to construct our novel classification methods on the hierarchical discipline structure.
% \subsection{Reviewer Assignment}
% The reviewer assignment is an old question occurrence with the development of the peer review system. Recent studies are rule-based and mainly focus on the conference paper's reviewer allocation~\cite{leyton2022matching}, e.g., how to both avoid COIs and balance each reviewer's workload. Some previous studies focus on scoring matches. 

%% file: document/conclusion.tex
\vspace{-0.5cm}
\section{Conclusion}
This paper proposed HIRPCN, a novel HMC method for the IRPC task on the real-world research proposal dataset. This method extracts the semantic information from each document separately and combines them to achieve a high-level abstraction. Then, the model process each previous level's prediction result and neighborhood topology structure on the pre-defined interdisciplinary graph as interdisciplinary knowledge. The prediction step integrates each part of the research proposal by attention mechanism and generates the next level prediction. The experiments show that our model achieves the best performance on three real-world datasets and can provide the best granular on each level prediction. Other experiments explains the attention mechanism and points out that our model could fix the incomplete interdisciplinary labels under a domain-specific expert evaluation. With the ability to start prediction from any given label, HIRPCN could assist whether researcher or fund administrator to fill the discipline information, which is crucial to improve the reviewer assignation.
% We first propose a two-level hierarchical encoder to capture the representation of the proposal fully. Then, we use a Transformer-based decoder incorporating the previous label prediction and the proposal's representation to generate the label. Due to the characteristic of the decoder, our model can start from the root label or any given previous label, which corresponds to utilize the expert knowledge. We conduct extensive experiments to evaluate our model performance. The results show that our model achieves the best performance in the HMPC problem and is effective for the utilization of expert knowledge. Other experiments evaluate the ability of our model to give a proper granularity and to capture the label dependency.

%% file: document/bio.tex
\vspace{-1.5cm}
\begin{IEEEbiography}[{\includegraphics[width=1in,height=1.25in,clip,keepaspectratio]{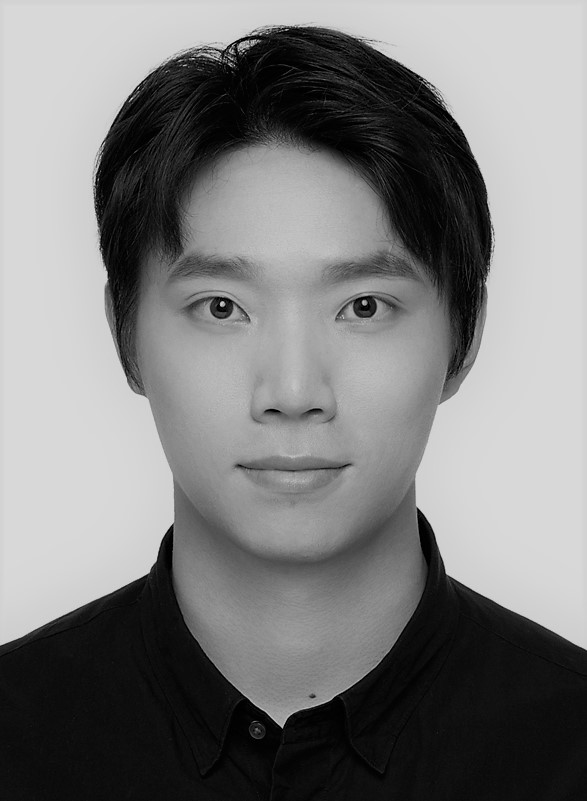}}]{Meng Xiao}
was born in 1995. He received his B.S. degree from the East China University of Technology and graduated from China University of Geosciences (Wuhan) with a master's degree. He is currently working toward obtaining a Ph.D. degree at the University of Chinese Academy of Sciences. His main research interests include Data Mining, Graph Representation Learning, Hierarchical Multi-label Text Classification combined with the knowledge, and the Knowledge Graphs of the discipline system.
\end{IEEEbiography}
\vspace{-1.5cm}
\begin{IEEEbiography}[{\includegraphics[width=1in,height=1.25in,clip,keepaspectratio]{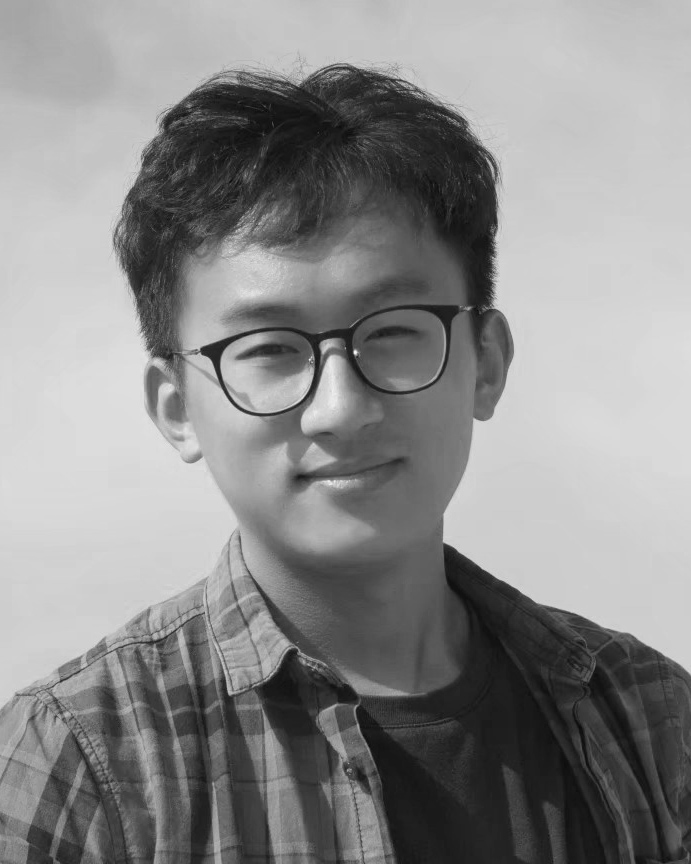}}]{Ziyue Qiao}
was born in 1996. He received his B.E. degree in 2017 from the Wuhan University, China. He is currently working toward obtaining a Ph.D. degree at the Computer Network Information Center, University of Chinese Academy of Sciences. His research interests include data mining, graph learning, and natural language processing, with an emphasis on designing new algorithms for graph representation/transfer learning and academic data mining.
\end{IEEEbiography}
\vspace{-1.5cm}
\begin{IEEEbiography}[{\includegraphics[width=1in,height=1.25in,clip,keepaspectratio]{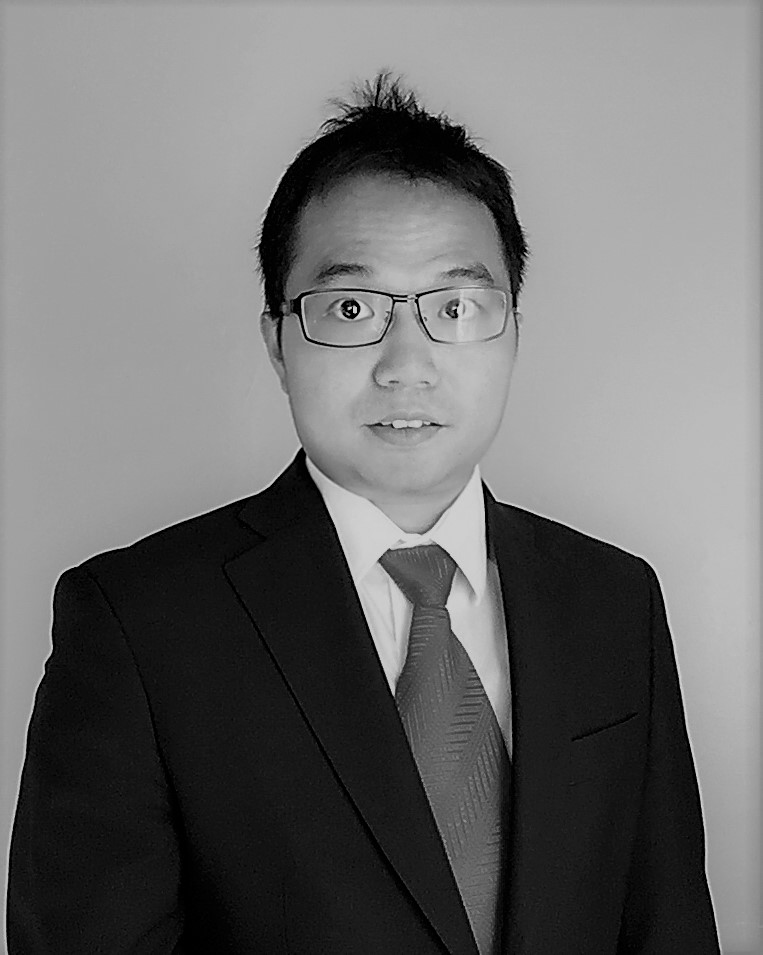}}]{Yanjie Fu}
is an assistant professor in the Department of Computer Science at the University of Central Florida. He received his Ph.D. degree from Rutgers, the State University of New Jersey in 2016, the B.E. degree from University of Science and Technology of China in 2008, and the M.E. degree from Chinese Academy of Sciences in 2011. His research interests include data mining and big data analytics.
\end{IEEEbiography}
\vspace{-1.5cm}
\begin{IEEEbiography}[{\includegraphics[width=1in,height=1.25in,clip,keepaspectratio]{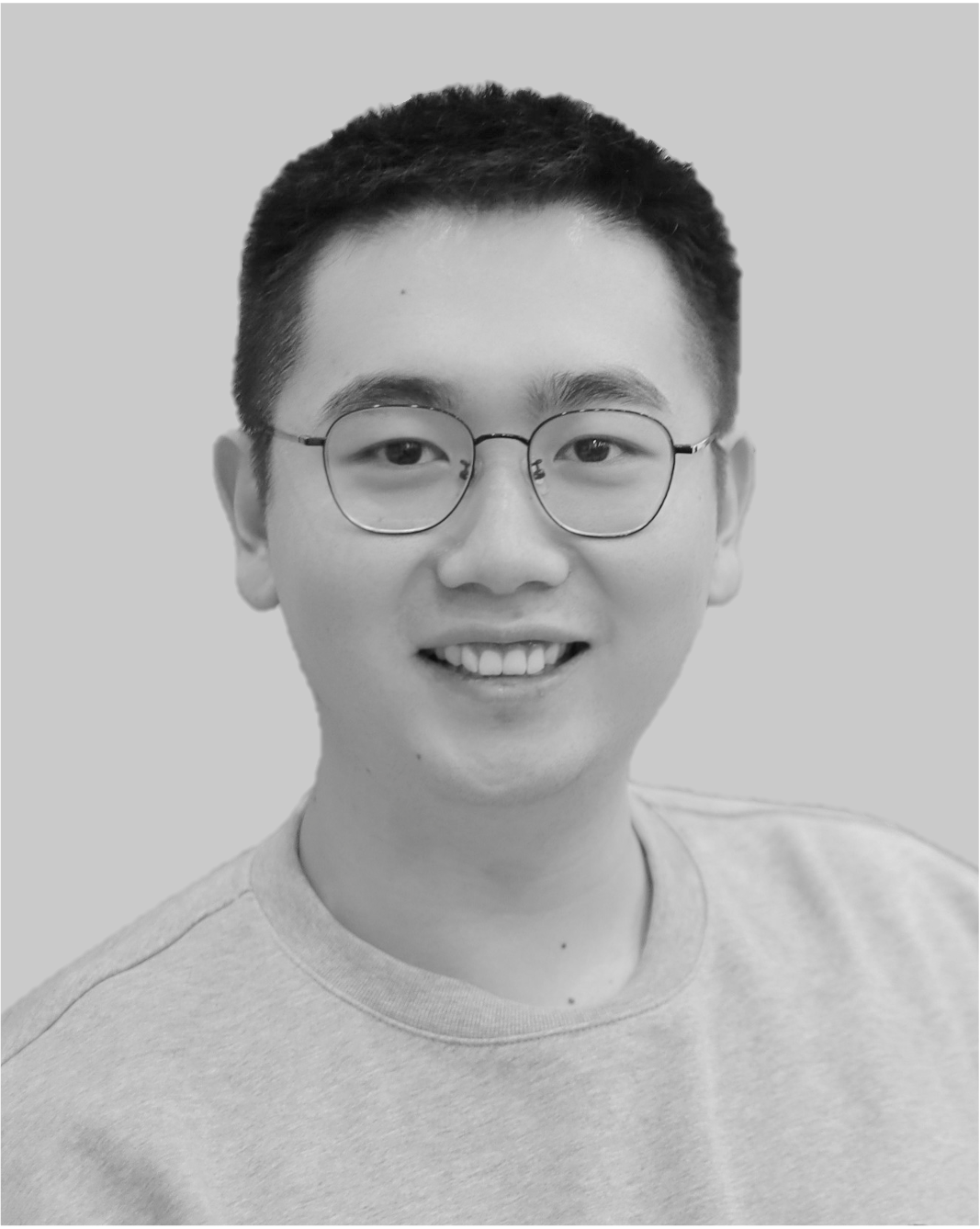}}]{Hao Dong} was born in 1997. He received his B.S. degree in 2019 from Shaanxi Normal University, China. He is currently working toward obtaining a Ph.D. degree at the Computer Network Information Center, University of Chinese Academy of Sciences. His research probes Data Mining, Natural Language Processing, Knowledge-Based Graph, and Temporal Graph.
\end{IEEEbiography}
\vspace{-1.5cm}
\begin{IEEEbiography}[{\includegraphics[width=1in,height=1.25in,clip,keepaspectratio]{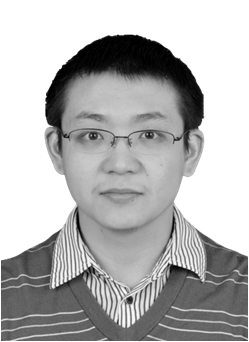}}]{Yi Du}
was born in 1988. He received his Ph.D. degree from Institute of Software, Chinese Academy of Sciences in 2013. He is a professor at the Department of Big Data Technology and Application Development at Computer Network Information Center, Chinese Academy of Sciences. His research interests include big data and visual analytics.
\end{IEEEbiography}
\vspace{-1.5cm}
\begin{IEEEbiography}[{\includegraphics[width=1in,height=1.25in,clip,keepaspectratio]{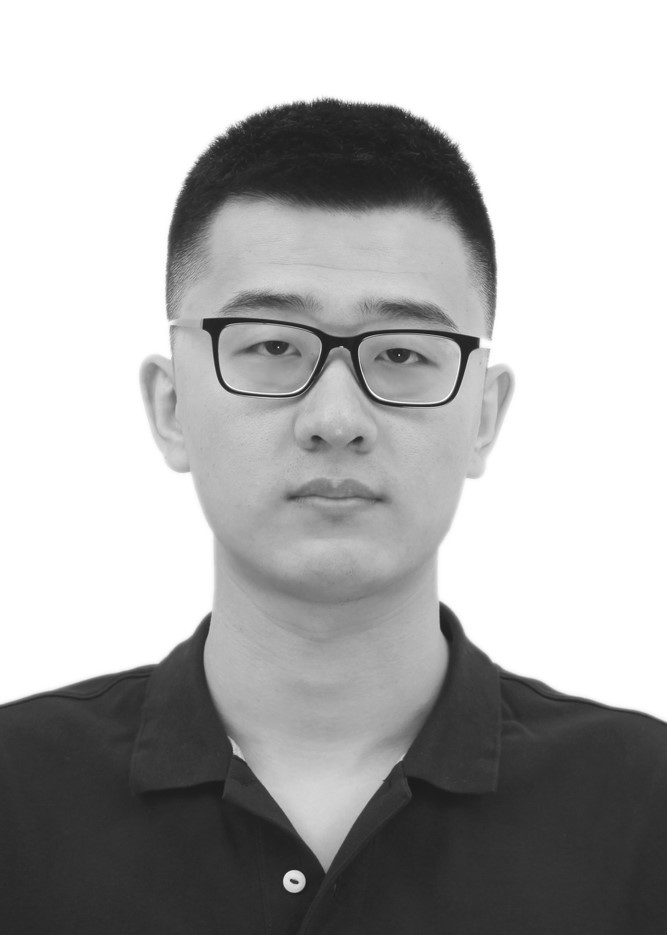}}]{Pengyang Wang}
is an assistant professor in the State Key Lab of Internet of Things and Smart Cities at the University of Macau. He received his Ph.D. in Computer Science from the University of Central Florida. He has broad interests in data mining and machine learning, especially in geospatial-temporal data modeling. He has published in related domains on top venues such as KDD, TKDE, IJCAI, WWW, AAAI etc. He also served as a (senior) program committee member among top conferences in the data mining and artificial intelligence communities.
\end{IEEEbiography}
\vspace{-1.5cm}
\begin{IEEEbiography}[{\includegraphics[width=1in,height=1.25in,clip,keepaspectratio]{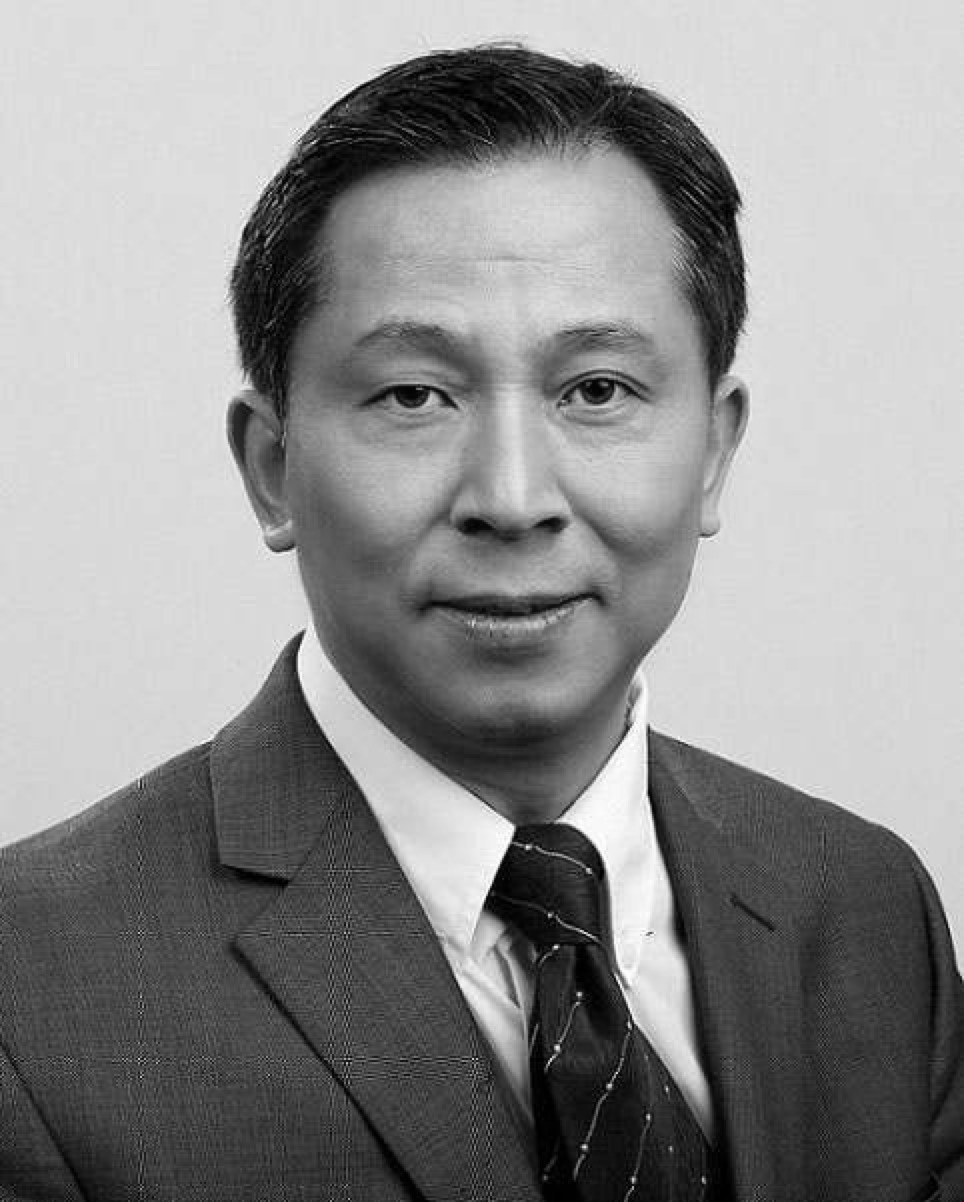}}]{Hui Xiong}
 is currently a Chair Professor at the
Hong Kong University of Science and Technol-
ogy (Guangzhou). Dr. Xiong’s research interests
include data mining, mobile computing, and their
applications in business. Dr. Xiong received his
PhD in Computer Science from University of
Minnesota, USA.
% He has served regularly on the
% organization and program committees of numer-
% ous conferences, including as a Program Co-
% Chair of the Industrial and Government Track for
% the 18th ACM SIGKDD International Conference
% on Knowledge Discovery and Data Mining (KDD), a Program Co-Chair
% for the IEEE 2013 International Conference on Data Mining (ICDM), a
% General Co-Chair for the 2015 IEEE International Conference on Data
% Mining (ICDM), and a Program Co-Chair of the Research Track for the
% 2018 ACM SIGKDD International Conference on Knowledge Discovery
% and Data Mining. He received the 2021 AAAI Best Paper Award and
% the 2011 IEEE ICDM Best Research Paper award. 
For his outstanding
contributions to data mining and mobile computing, he was elected an
AAAS Fellow and an IEEE Fellow in 2020.
\end{IEEEbiography}
\vspace{-1.5cm}
\begin{IEEEbiography}[{\includegraphics[width=1in,height=1.25in,clip,keepaspectratio]{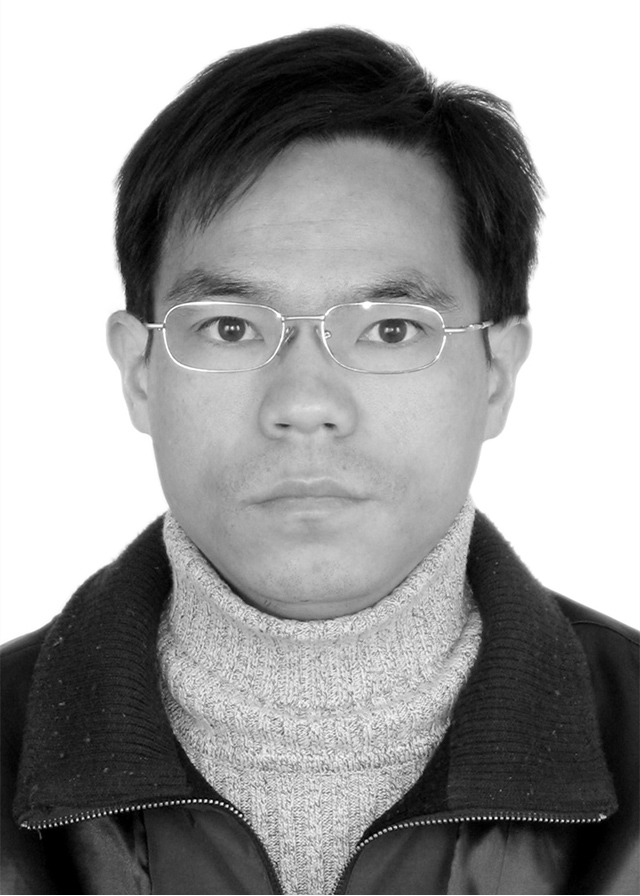}}]{Yuanchun Zhou}
was born in 1975. He received his Ph.D. degree from Institute of Computing Technology, Chinese Academy of Sciences, in 2006. He is a professor, Ph.D. supervisor, the deputy director of  Computer Network Information Center, Chinese Academy of Sciences. His research interests include data mining, big data processing, and knowledge graph.
\end{IEEEbiography}